\documentclass[english,aps, numerical, nofootinbib]{revtex4}
\usepackage[T1]{fontenc}
\usepackage[latin9]{inputenc}
\usepackage{amsmath}
\usepackage{amssymb}
\usepackage{esint}

\makeatletter
\@ifundefined{textcolor}{}
{%
 \definecolor{BLACK}{gray}{0}
 \definecolor{WHITE}{gray}{1}
 \definecolor{RED}{rgb}{1,0,0}
 \definecolor{GREEN}{rgb}{0,1,0}
 \definecolor{BLUE}{rgb}{0,0,1}
 \definecolor{CYAN}{cmyk}{1,0,0,0}
 \definecolor{MAGENTA}{cmyk}{0,1,0,0}
 \definecolor{YELLOW}{cmyk}{0,0,1,0}
 }

\makeatother

\usepackage{babel}

\begin{document}
\global\long\def\f#1{\mathrm{f}^{#1}}
 \global\long\def\g{g_{\mathrm{c}}}
 \global\long\def\ge{g_{\mathrm{end}}}
 \global\long\def\gf{\frac{\g^{2}}{f}}
 \global\long\def\fgh{g_{\mathrm{c}}^{2}/f}
 \global\long\def\w#1#2{\delta W_{#1}^{#2}}
 \global\long\def\k#1{\mathbf{k}_{#1}}
 \global\long\def\te#1#2{T_{#1}^{\mathrm{E}}\left(\hat{\mathbf{k}}_{#2}\right)}
 \global\long\def\mpl{m_{\mathrm{Pl}}}
 \global\long\def\m#1#2#3{\mathcal{M}_{#2}^{#1}\left(\hat{\mathbf{k}}_{#3}\right)}
 \global\long\def\gfi{\frac{\g^{2}}{f_{0}}}
 \global\long\def\fnl{f_{\mathrm{NL}}}
 \global\long\def\gz{g_{\zeta}}
 \global\long\def\gcl{g_{3}^{\mathrm{cl}}}
 \global\long\def\jc{\varphi_{\mathrm{c}}}
 \global\long\def\ze{\zeta_{\mathrm{e}}}
 \global\long\def\e{\mathrm{e}}

\title{The Primordial Curvature Perturbation from Vector Fields of General
non-Abelian Groups}

\author{Mindaugas Kar\v{c}iauskas}

\affiliation{CAFPE and Departamento de F\'isica Te\'orica y del Cosmos, Universidad
de Granada, Granada-18071, Spain}

\email{mindaugas@ugr.es}
\begin{abstract}
We consider the generation of primordial curvature perturbation by
general non-Abelian vector fields without committing to a particular
group. Self-interactions of non-Abelian fields make the field perturbation
non-Gaussian. We calculate the bispectrum of the field perturbation
using the in-in formalism at tree level. The bispectrum is dominated
by the classical evolution of fields outside the horizon. In view
of this we show that the dominant contribution can be obtained from
the homogeneous classical equation of motion. Then we calculate the
power spectrum of the curvature perturbation. The anisotropy in spectrum
is suppressed by the number of fields. This makes it possible for
vector fields to be responsible for the total curvature perturbation
in the Universe without violating observational bounds on statistical
anisotropy. The bispectrum of the curvature perturbation is also anisotropic.
Finally we give an example of the end-of-inflation scenario in which
the curvature perturbation is generated by vector gauge fields through
varying gauge coupling constant(s), which in covariant derivatives
couples the Higgs field to the vector fields. We find that reasonably
large gauge groups may result in the observable anisotropy in the
power spectrum of the curvature perturbation.
\end{abstract}
\maketitle

\section{Introduction}

Inflation was proposed to alleviate the horizon and flatness problems
of the Hot Big Bang cosmology \citep{Starobinsky(1980),Guth(1980)infl}.
Currently it still is arguably the most compelling mechanism to explain
the high degree of homogeneity and isotropy of the Universe as well
as its flatness. The case for inflation was further strengthened after
the release of first detailed observations of the Cosmic Microwave
Background (CMB) radiation. The spectrum of the temperature perturbation
of the CMB was found to be consistent with inflationary predictions
and ruled out the rival theory of cosmic strings as the primary origin
of this perturbation. Moreover, with the increasing precision of observational
data it is becoming possible not only to falsify the competing theories
of the origin of the primordial perturbation, but to falsify different
models of inflation themselves. Measurements of temperature irregularities
in the CMB sky provide a powerful tool to probe the physics of the
very early Universe and the increasing precision allows to do it in
a more and more detail.

The simplest inflationary models, with a scalar field driving inflation
and producing the primordial curvature perturbation $\zeta$, are
still consistent with current CMB data. But several tentative anomalies,
which are persistently found in all WMAP data releases, might suggest
a need of more complex models. For example, the observed power asymmetry
\citep{Hansen_etal(2008)powerAsym,Eriksen_etal(2004)}, the alignment
of low-$l$ CMB multipoles \citep{Tegmark_etal(2003)axisOfEvil,Bielewicz_etal(2004)axisOfEvil,Land_Magueijo(2005)_AxisOfEvil,Copi_etal(2005)axisOfEvil}
or a deep cold spot in the southern Galactic hemisphere \citep{Cruz_etal(2006)coldSpt,Cruz_etal(2007)coldSpt}.
If the origin of these anomalies is confirmed to be primordial it
will imply some degree of the statistical inhomogeneity and/or anisotropy
of the primordial curvature perturbation. Such anomalies cannot be
explained by the simplest inflationary models invoking only scalar
fields. Scalar fields do not generate statistical anisotropy as they
do not choose a preferred direction. But vector fields do, and if
statistical anisotropy is established to be of primordial origin,
the most natural way to explain it is with effects of vector fields.

There are two ways through which non-negligible contribution of vector
fields to the evolution of the universe can induce statistical anisotropy.
First, if a vector field have an effect on the global expansion of
the universe, the latter will be anisotropic. The anisotropic expansion
during inflation causes statistically anisotropic quantum fluctuations
of light scalar fields. When these fluctuations become classical and
cause perturbations in the metric, the latter are statistically anisotropic
too as well as the temperature irregularities of CMB. This mechanism
was first considered in \citep{Ackerman_etal(2007)}. On the other
hand, the energy density of vector fields can be negligible, so that
the global expansion is approximately isotropic, but perturbations
of vector fields themselves generate or contribute to $\zeta$. In
general quantum fluctuations of vector fields are statistically anisotropic
making their contribution to $\zeta$ statistically anisotropic too.
The generation of $\zeta$ by vector fields was first considered in
Ref.~\citep{Dimopoulos2006}. However, this and several subsequent
papers \citep{Dimopoulos2007,Dimopoulos_Karciauskas(2008)} did not
consider statistical anisotropy, which was first considered in Ref.~\citep{Yokoyama_Soda(2008)}.
The comprehensive study of statistical anisotropy generated by a vector
field can be found in Ref.~\citep{Dimopoulos_etal_anisotropy(2008)}
(see also Ref.~\citep{Karciauskas_etal(2008)} for computations of
anisotropic non-linearity parameter $\fnl$).

Recently the interest in statistical anisotropy and vector fields
has grown considerably. Authors of Refs.~\citep{Toledo_etal(2009),Toledo_etal(2010),Toledo_etal(2011)Feynman,Bartolo_etal(2009)_Bispectrum,Bartolo_etal(2009)Trispectrum,Dimastrogiovanni_etal(2010)NonAbelianReview}
studied two-, three- and four-point correlators of the curvature perturbation
generated by vector fields in detail. While in Refs.~\citep{Dimopoulos_us(2009)_fF2,Dimopoulos_us(2009)_fF2_PRL}
a model with a massive vector curvaton field is presented in which
the vector field can generate both statistically isotropic and anisotropic
perturbation. In these works a negligible contribution of vector fields
to the global expansion rate is assumed. 

Another line of research was concentrated on effects of anisotropic
inflation on the curvature perturbation \citep{Watanabe_etal(2010)anisoBckgr,Dulaney_etal(2010)anisoBckgr,Gumrukcuoglu_etal(2010)anisoCMB}.
The universe during inflation expands anisotropically if the backreaction
of the vector field is non-negligible. Such setup is considered in
Refs.~\citep{Ackerman_etal(2007),Watanabe_Kanno_Soda(2009),Kanno_Soda_Watanabe(2009)PMFs,Emami_etal(2010)backreaction,Kanno_etal(2010)attractor,Dimopoulos_Wagstaff(2010)BackReact,Hervik_etal(2011)BIattract}.
Particularly interesting are the results of Refs.~\citep{Kanno_etal(2010)attractor,Dimopoulos_Wagstaff(2010)BackReact}.
Authors of these papers considered a model with time varying kinetic
function of an Abelian vector field of the form $f\left(t\right)F_{\mu\nu}F^{\mu\nu}$,
where $F_{\mu\nu}$ is the field strength tensor $F_{\mu\nu}=\partial_{\mu}A_{\nu}-\partial_{\nu}A_{\mu}$.
It was shown that if $f\left(t\right)$ is modulated by the inflaton,
then the scaling of the form $f\propto a^{-4}$ ($a$ being a scale
factor) is an attractor solution for a large parameter space. This
is very significant as such scaling leads to the flat perturbation
spectrum for the vector field \citep{Dimopoulos_us(2009)_fF2,Dimopoulos_us(2009)_fF2_PRL}.
$fF^{2}$ coupling also induces anisotropy in inflationary expansion
of the order of the slow-roll parameter. In addition in Ref.~\citep{Dimopoulos_Wagstaff(2010)BackReact}
it was shown that the vector field backreaction slows down the inflaton,
i.e. the inflaton potential is effectively \textquotedbl{}flattened\textquotedbl{}.
The backreaction of non-Abelian vector fields with the time-varying
kinetic function was also considered in Ref.~\citep{Murata_Soda(2011)nAb-attractor}.

The interest in detecting statistical anisotropy in the CMB is increasing
too \citep{Pullen_Kamionkowski(2007),Groeneboom&Eriksen_CMBanisotropy(2009),Hanson_Lewis(2009)CMBanisotropy,Groeneboom_etal(2009)anisotropy2,Hanson_etal(2010)statAnis,Ma_etal(2011)statAnis}.
The anisotropy in the power spectrum of the primordial curvature perturbation
can be parametrized as\begin{equation}
\mathcal{P}_{\zeta}\left(k\right)=\mathcal{P}_{\zeta}^{\mathrm{iso}}\left(k\right)\left[1+g_{\zeta}\left(k\right)\left(\hat{\mathbf{n}}\cdot\hat{\mathbf{k}}\right)^{2}\right],\label{eq:Pz-aniso-def}\end{equation}
where only a quadrupole term is kept, $\hat{\mathbf{n}}$ and $\hat{\mathbf{k}}$
are unit vectors and the amplitude $g_{\zeta}\left(k\right)$ is in
general a function of wavenumber $k$. First results of measuring
$g_{\zeta}$ were given in Ref.~\citep{Groeneboom&Eriksen_CMBanisotropy(2009)}.
After correcting a mistake in this work Refs.~\citep{Hanson_Lewis(2009)CMBanisotropy,Groeneboom_etal(2009)anisotropy2}
published consistent results detecting the departure from statistical
isotropy at very high $9\sigma$ significance level $g_{\zeta}=0.29\pm0.031$.
They also found that the preferred direction $\hat{\mathbf{n}}$ is
very close to the ecliptic pole $\left(l,b\right)=\left(96,30\right)$.
The proximity of $\hat{\mathbf{n}}$ to the ecliptic pole suggests
very strongly that the detected $g_{\zeta}$ is due to a systematic
effect. This was indeed discussed in Refs.~\citep{Hanson_etal(2010)statAnis,Groeneboom_etal(2009)anisotropy2}
were they investigated whether a non-zero $g_{\zeta}$ could be accounted
for by the WMAP beam asymmetry or other systematics. Unfortunately,
both works reached somewhat contradicting conclusions. The authors
of Ref.~\citep{Groeneboom&Eriksen_CMBanisotropy(2009)} could not
determine a systematic effect causing such a large $g_{\zeta}$, while
Ref.~\citep{Hanson_etal(2010)statAnis} claims it is due to the beam
asymmetry and give the bound $\left|g_{\zeta}\right|<0.07$. Despite
this, as the precision of measurements increase, the prospect of detecting
statistical anisotropy of primordial origin is very exciting. Such
anisotropy offers a new observable to probe the physics of the very
early Universe. Indeed, as shown in Refs.~\citep{Pullen_Kamionkowski(2007),Ma_etal(2011)statAnis}
with the Planck temperature data alone it will be possible to constrain
$g_{\zeta}$ with an accuracy of $0.01$ $\left(2\sigma\right)$.
In addition, the polarization data alone will offer $0.03$ accuracy
as the consistency check. Furthermore, an extended Planck mission
can constrain the spectral index of $g_{\zeta}\propto k^{q}$ to an
accuracy of $\Delta q\sim0.3$ $\left(1\sigma\right)$.

The interest in measuring the anisotropy of higher order correlators
is increasing too \citep{Rudjord_etal(2009)anisotropicfNL,Shiraishi_Yokoyama(2011)statanisCMB,Bartolo_etal(2011)anis_estim}.
This is justified, because higher order correlators can be predominantly
anisotropic as shown in Refs.~\citep{Yokoyama_Soda(2008),Dimopoulos_etal_anisotropy(2008),Karciauskas_etal(2008),Dimopoulos_us(2009)_fF2,Dimopoulos_us(2009)_fF2_PRL,Toledo_etal(2009),Toledo_etal(2010),Bartolo_etal(2009)_Bispectrum,Bartolo_etal(2009)Trispectrum}.
This was specifically emphasized in Refs.~\citep{Karciauskas_etal(2008),Dimopoulos_us(2009)_fF2,Dimopoulos_us(2009)_fF2_PRL,Karciauskas_thesis}
were non-linearity parameter $\fnl$ was calculated for several models.
It was found that even if anisotropy in the power spectrum is subdominant,
$\fnl$ can be predominantly anisotropic with the same preferred direction
$\hat{\mathbf{n}}$ as the spectrum. Moreover, the magnitude of $\fnl$
is proportional to anisotropy in the spectrum $g_{\zeta}$. If such
non-Gaussianity is detected it would be a smoking gun for the vector
field contribution to the primordial curvature perturbation. Impacts
of statistically anisotropic primordial perturbation on CMB observables
were studied in Refs.~\citep{Gumrukcuoglu_etal(2010)anisoCMB,Watanabe_etal(2010)anisoCMB}.

Instead of vector fields being responsible only for the curvature
perturbation, Refs.~\citep{Ford(1989)_Vtriad,Golovnev_etal(2008),Golovnev_etal(2009),Golovnev_etal(2008)GWs,Maleknejad_etal(2011)short,Maleknejad_etal(2011)long}
also propose models in which vector fields drive inflation. Such models
face a challenge of making expansion of the universe predominantly
isotropic. The two suggestions proposed are either introducing an
orthogonal triad of vector fields or a large number of randomly oriented
ones. Particularly, authors of Refs.~\citep{Maleknejad_etal(2011)short,Maleknejad_etal(2011)long}
also consider non-Abelian vector fields. Their setup consists of an
orthogonal triad of equal norm vector fields of $SU\left(2\right)$
group. In this case the total energy-momentum tensor of vector fields
is isotropic and if they dominate the universe, the latter inflates
isotropically.

The study of vector fields in inflationary cosmology are not only
interesting from the phenomenological point of view, as means of explaining
CMB anomalies. Such models are also interesting from the theoretical
perspective. The possibility of vector fields affecting or generating
the total curvature perturbation in the Universe opens a new window
for inflationary model building. We might no longer need direct involvement
of scalars to create $\zeta$; it could be created by vector fields
with, for example, varying couplings. In particle physics models non-Abelian
vector fields are much more common than Abelian ones. In addition,
as large non-Abelian gauge groups have many vector fields, assuming
their random orientation, it is natural to expect the suppression
of statistical anisotropy of $\zeta$.

In this paper we study the generation of $\zeta$ by non-Abelian vector
fields. The effects of $SU\left(2\right)$ non-Abelian vector fields
on the curvature perturbation were first studied in Refs.~\citep{Bartolo_etal(2009)_Bispectrum,Bartolo_etal(2009)Trispectrum,Dimastrogiovanni_etal(2010)NonAbelianReview}.
In this paper we demonstrate that non-Gaussian correlators are dominated
by the interactions outside horizon. First, we make use of the full
quantum formalism, the so called {}``in-in formalism'', to calculate
the bispectrum at tree level and show that it is dominated by the
classical part. That is, by interactions of fields after horizon crossing.
Since this is the case, correlation functions can be calculated much
easier using only the homogeneous equation of motion of vector fields.
We perform this calculation explicitly and show that the result obtained
is the same as the dominant part of the result calculated in the in-in
formalism.

We also calculate the spectrum and bispectrum of the curvature perturbation.
It is found that the anisotropy in the spectrum of $\zeta$ is suppressed
by the number of vector fields involved in generating $\zeta$ (assuming
these are oriented randomly). Thus with a large enough gauge group
$\zeta$ can be generated solely from vector fields without violating
observational bound on the statistical anisotropy. However, it is
possible that a small detectable anisotropy remains.

The bispectrum of $\zeta$ from non-Abelian vector fields is also
anisotropic. Although the form of anisotropy is more complicated than
in the single field case as it involves not one but many preferred
directions.

In the last section we give a simple example of a non-Abelian gauge
field generating $\zeta$. We use the end-of-inflation scenario in
which $\zeta$ is generated by the gauge fields through varying gauge
coupling constant(s) in the covariant derivative, which couples Higgs
field to the gauge fields.

The paper is organized as follows. In section~\ref{sec:Lagrangian-and-Spectrum}
we define the Lagrangian and the setup of our model. In section~\ref{sec:quantum}
the bispectrum at tree level is calculated using the quantum in-in
formalism. In section~\ref{sec:classical} it is shown that the classical
calculation of the bispectrum from the homogeneous equation of motion
gives exactly the same result as the dominant part of the bispectrum
in the previous section. Using the $\delta N$ formula, the correlators
of $\zeta$ are calculated in section~\ref{sec:curvature-perturbation}.
Finally, an example of a mechanism for the generation of $\zeta$
from non-Abelian gauge fields is given in section~\ref{sec:Example}.
It is generated through the gauge couplings of the Higgs field. A
summary is presented in section~\ref{sec:conclusions}.

In this paper we use natural units, where $c=\hbar=1$ and Newton's
gravitational constant is $8\pi G=\mpl^{-2}$, where $\mpl$ is the
reduced Planck mass.

\section{The Lagrangian\label{sec:Lagrangian-and-Spectrum}}

Consider a general Lagrangian of non-Abelian vector fields\begin{equation}
\mathcal{L}=-\frac{1}{4g^{2}}F_{\mu\nu}^{a}F_{a}^{\mu\nu},\label{eq:Lagrangian-def}\end{equation}
where the field strength tensor $F_{\mu\nu}^{a}$ is\begin{equation}
F_{\mu\nu}^{a}=\partial_{\mu}A_{\nu}^{a}-\partial_{\nu}A_{\mu}^{a}+\f{abc}A_{\mu}^{b}A_{\nu}^{c}.\label{eq:F-def}\end{equation}
We do not specify the gauge group, $\f{abc}$ are structure constants
of the Lie algebra of any non-Abelian group and they are antisymmetric
in permutations of indices $a,\, b$ and $c$. 

In this paper we are interested in a time varying $g$ but we do not
specify the origin of this variation. The modulation of $g$ might
be due to some scalar degree of freedom. The inflaton itself can modulate
$g$ (for such models with Abelian vector field see Refs.~\citep{Watanabe_Kanno_Soda(2009),Emami_etal(2010)backreaction,Kanno_etal(2010)attractor,Dimopoulos_Wagstaff(2010)BackReact}
and non-Abelian ones see Ref.~\citep{Murata_Soda(2011)nAb-attractor}),
but in this paper we do not need to specify the origin of time dependence
of $g$. 

Let us recast the Lagrangian in Eq.~\eqref{eq:Lagrangian-def} by
factoring $g\left(t\right)$ into a constant and time dependent parts
in the following way \begin{equation}
g\left(t\right)\equiv\g/\sqrt{f\left(t\right)}.\label{eq:gc-def}\end{equation}
$\g$ is the constant value of $g$ when the modulating degree of
freedom is stabilized at time $t_{\mathrm{s}}$. At this moment the
kinetic function $f\left(t_{\mathrm{s}}\right)=1$. The constant $\g$
can be absorbed into the field strength tensor by the field redefinition
$\tilde{A}_{\mu}^{a}\equiv A_{\mu}^{a}/\g$. Dropping tildes the Lagrangian
in Eq.~\eqref{eq:Lagrangian-def} then becomes\begin{equation}
\mathcal{L}=-\frac{1}{4}f\, F_{\mu\nu}^{a}F_{a}^{\mu\nu},\label{eq:Lagrangian}\end{equation}
where the redefined field strength tensor becomes \begin{equation}
F_{\mu\nu}^{a}=\partial_{\mu}A_{\nu}^{a}-\partial_{\nu}A_{\mu}^{a}+\g\f{abc}A_{\mu}^{b}A_{\nu}^{c}\label{eq:F-def2}\end{equation}
with $\g$ being a self-coupling constant.

Using the gauge freedom of the non-Abelian vector field in Eq.~\eqref{eq:Lagrangian}
we can always choose a gauge in which any given space-time component
of all the vector fields is zero \citep{Weinberg_book2}. However,
in general it is not possible to make all four components of all vector
fields in a group vanish by a gauge choice. This is in contrast to
the Abelian field, where one can always choose a gauge in which the
classical or homogeneous part of the field is zero. For the rest of
the paper let us choose the vanishing component to be the temporal
one so that $A_{0}^{a}=0$ for all $a$. For the remainder of the
paper we use the convention of space indices denoted by subscripts
and gauge ones by superscripts except in those cases where equations
are written in Lorentz covariant four-vector form.

The energy-momentum tensor of the massless vector field is anisotropic
\citep{Dimopoulos2006}. If the energy density of such a field is
non-negligible the expansion of the universe becomes anisotropic.
To avoid excessive large scale anisotropy one can introduce a large
number of vector fields, suppressing the anisotropy by a factor of
$\sqrt{\mathcal{N}}$, where $\mathcal{N}$ is the number of vector
fields. Such mechanism is employed in vector inflation \citep{Golovnev_etal(2008)}.
In our setup this could be achieved by taking a very large gauge group.
Another possibility is to introduce three identical, orthogonal vector
fields \citep{ArmendarizPicon(2004)triad,Golovnev_etal(2008)}. This
option was recently explored with $SU\left(2\right)$ vector fields
in Refs.~\citep{Maleknejad_etal(2011)long,Maleknejad_etal(2011)short}.
In this paper we consider a third possibility, namely we assume that
the energy density of the vector field is negligible during inflation.
In other words, we neglect a backreaction of the vector field on the
expansion of the universe. However vector fields can still generate
the curvature perturbation. This might happen, for example, in the
vector curvaton \citep{Dimopoulos2006} or end-of-inflation \citep{Yokoyama_Soda(2008)}
scenarios. In the former case, the non-Abelian vector fields must
acquire a mass through a Higgs mechanism prior to generating $\zeta$.
An example of the latter case will be given in section~\ref{sec:Example}.

Taking the contribution of the vector field to the total energy density
to be negligible, inflation can be assumed to be isotropic and we
can use the Friedmann-Lemaître-Robertson-Walker (FLRW) background
with the metric $g_{\mu\nu}=\mathrm{diag}\left[1,-a^{2}\left(t\right),-a^{2}\left(t\right),-a^{2}\left(t\right)\right]$,
where $t$ is the cosmic time. We will also use the conformal time
$\tau\equiv\int\mathrm{d}t/a\left(t\right)$ in the paper. 

Expanding the Lagrangian in Eq.~\eqref{eq:Lagrangian} in the FLRW
background we find a term\begin{equation}
\mathcal{L}\supset a^{-4}\g^{2}\f{abc}\f{ade}A_{i}^{b}A_{j}^{c}A_{i}^{d}A_{j}^{e},\label{eq:L-part1}\end{equation}
in which the scale factor appears explicitly. But the normalization
of the scale factor is arbitrary while the Lagrangian is a physical
quantity related to the energy of the system and cannot contain arbitrary
normalizable factors. The appearance of $a$ in Eq.~\eqref{eq:L-part1}
is due to the fact that the vector field $A_{i}^{a}$ is defined with
respect to the comoving coordinates. The physical vector field, defined
with respect to the physical coordinates is $A_{i}^{a}/a$ \citep{Dimopoulos2006,Dimopoulos_etal_anisotropy(2008),Karciauskas_thesis}.
Therefore, it will be useful to define a physical, canonically normalized
vector field\begin{equation}
W_{i}^{a}=\sqrt{f}\frac{A_{i}^{a}}{a}.\label{eq:W-def}\end{equation}
We also use Fourier modes $\delta W_{i}^{a}\left(\mathbf{k}\right)$
of the perturbation of $W_{i}^{a}$ \begin{equation}
\delta W_{i}^{a}\left(\mathbf{x}\right)=\int\frac{\mathrm{d}^{3}k}{\left(2\pi\right)^{3}}\delta W_{i}^{a}\left(\mathbf{k}\right)\mathrm{e}^{i\mathbf{k}\cdot\mathbf{x}}.\label{eq:W-Fourier}\end{equation}

\section{Correlation Functions in the in-in Formalism\label{sec:quantum}}

\subsection{The in-in Formalism}

In this section we calculate the three point correlation function
of vector field perturbation. Following Refs.~\citep{Maldacena(2002),Weinberg(2005)nG}
we use the so called {}``in-in formalism''. In our setup, the calculation
is simplified due to the assumption that the energy density of vector
fields during inflation is negligible, which means that it is not
necessary to consider the corresponding perturbation of the metric.

In the in-in formalism expectation values $\left\langle 0\right|\hat{O}\left|0\right\rangle $
are calculated in the interaction picture. In this picture both, field
operators and state vectors, varies with time. The variation of the
former is governed by the equation of motion of the free field, while
the latter evolves due to interaction terms. To find free field operators
for non-Abelian vector fields we note that each massless vector field
has two degrees of freedom. The field is quantized by promoting these
degrees of freedom to operators with canonical commutation relations.
Thus $\delta W_{i}^{a}\left(\mathbf{k}\right)$ in Eq.~\eqref{eq:W-Fourier}
becomes\begin{equation}
\delta\hat{W}_{i}^{a}\left(\mathbf{k},\tau\right)=\sum_{\lambda=\mathrm{L},\mathrm{R}}\left[e_{i}^{\lambda}\left(\hat{\mathbf{k}}\right)w\left(k,\tau\right)\hat{a}_{\lambda}^{a}\left(\mathbf{k}\right)-e_{i}^{\lambda*}\left(-\hat{\mathbf{k}}\right)w^{*}\left(k,\tau\right)\hat{a}_{\lambda}^{a\dagger}\left(-\mathbf{k}\right)\right],\label{eq:W-quantization}\end{equation}
where $\tau$ is the conformal time. The raising and lowering operators
in this equation satisfy canonical commutation relations\begin{equation}
\left[a_{\lambda}^{a}\left(\mathbf{k}\right),a_{\lambda'}^{b\dagger}\left(-\mathbf{k}'\right)\right]=\left(2\pi\right)^{3}\delta\left(\mathbf{k}+\mathbf{k}'\right)\delta_{\lambda\lambda'}\delta_{ab}\end{equation}
with others being zero, while the left- and right-handed circular
polarization vectors $e_{i}^{\lambda}$ are chosen in such a way that
$e_{i}^{\lambda}\left(-\hat{\mathbf{k}}\right)=-e_{i}^{\lambda*}\left(\hat{\mathbf{k}}\right)$.
With $\hat{\mathbf{k}}=\left(0,0,k\right)$ they become\begin{equation}
e_{i}^{\mathrm{L}}\left(\hat{\mathbf{k}}\right)=\frac{1}{\sqrt{2}}\left(1,i,0\right)\quad\mathrm{and}\quad e_{i}^{\mathrm{R}}\left(\hat{\mathbf{k}}\right)=\frac{1}{\sqrt{2}}\left(1,-i,0\right).\end{equation}
In the interaction picture $\delta\hat{W}_{i}^{a}\left(\mathbf{k},\tau\right)$
is a free quantum field. Thus the conjugate pair $\left\{ w,w^{*}\right\} $
in Eq.~\eqref{eq:W-quantization} are solutions of a free vector
field equation of motion. In principle $\left\{ w,w^{*}\right\} $
should have polarization and group indices, e.g. $w_{\lambda}^{a}$.
However, the Lagrangian in Eq.~\eqref{eq:Lagrangian} does not have
parity violating terms and both polarization modes satisfy the same
equation of motion. The same is true for the gauge index, all vector
fields satisfy the same free field equation of motion with the same
initial conditions. Thus to simplify the notation we dropped polarization
and gauge indices out.

In Refs.~\citep{Dimopoulos_us(2009)_fF2,Dimopoulos_us(2009)_fF2_PRL}
it was found that $w$ satisfies the same equation of motion as the
scalar field if the kinetic function during inflation varies with
time as $f\propto a^{-1\pm3}$. For a massless field in de Sitter
background this equation becomes\begin{equation}
\ddot{w}+3H\dot{w}+\left(\frac{k}{a}\right)^{2}w=0,\end{equation}
where dots denote derivatives with respect to the cosmic time $t$.
The solution of this equation is very well known. With Bunch-Davies
vacuum initial conditions it is given in conformal time by\begin{equation}
w=\frac{H}{\sqrt{2k^{3}}}\left(1-ik\tau\right)\mathrm{e}^{-ik\tau}.\end{equation}

With these definitions the Wightman function becomes\begin{equation}
\left\langle 0\left|\delta\hat{W}_{i}^{a}\left(\mathbf{k},\tau\right)\delta\hat{W}_{j}^{b}\left(\mathbf{k}',\tau'\right)\right|0\right\rangle =\left(2\pi\right)^{3}\delta\left(\mathbf{k}+\mathbf{k}'\right)\delta_{ab}\te{ij}{}w\left(k,\tau\right)w^{*}\left(k,\tau'\right),\label{eq:2pt-corr}\end{equation}
where the tensor $\te{ij}{}$ is defined by \citep{Dimopoulos_etal_anisotropy(2008),Karciauskas_thesis}\begin{equation}
\te{ij}{}\equiv e_{i}^{L}\left(\hat{\mathbf{k}}\right)e_{j}^{R}\left(\hat{\mathbf{k}}\right)+e_{i}^{\mathrm{R}}\left(\hat{\mathbf{k}}\right)e_{j}^{\mathrm{L}}\left(\hat{\mathbf{k}}\right)=\delta_{ij}-\hat{k}_{i}\hat{k}_{j}.\label{eq:T-def}\end{equation}

N-point correlation functions of the vector field perturbation are
calculated in the interaction picture as vacuum expectation values
of the form \begin{equation}
g_{N}\left(\mathbf{x}_{1},\mathbf{x}_{2},\ldots,\mathbf{x}_{N}\right)=\left\langle 0\left|\hat{U}^{-1}\delta\hat{W}_{l}^{f}\left(\mathbf{x}_{1},\tau\right)\delta\hat{W}_{m}^{g}\left(\mathbf{x}_{2},\tau\right)\ldots\delta\hat{W}_{n}^{h}\left(\mathbf{x}_{N},\tau\right)\hat{U}\right|0\right\rangle ,\label{eq:g-def}\end{equation}
where, to simplify notation, we also suppressed gauge and space indices
in the function $g_{N}$. The unitary operator $\hat{U}$ is given
by\begin{equation}
\hat{U}=\exp\left\{ -i\int_{\tau_{0}}^{\tau}\hat{H}_{\mathrm{int}}\left(\tau'\right)\mathrm{d}\tau'\right\} \label{eq:U-def}\end{equation}
with $\tau_{0}$ being some early time when the mode of interest is
deep within the horizon. The interaction Hamiltonian $\hat{H}_{\mathrm{int}}$
can be found from the interaction terms of the Lagrangian in Eq.~\eqref{eq:Lagrangian}.
With the temporal gauge ($A_{0}^{a}=0$) these terms become\begin{equation}
\mathcal{L}_{\mathrm{int}}=-a^{-4}\frac{1}{2}f\left[\g\f{abc}\left(\partial_{i}A_{j}^{a}-\partial_{j}A_{i}^{a}\right)A_{i}^{b}A_{j}^{c}+\frac{1}{2}\g^{2}\f{abc}\f{ade}A_{i}^{b}A_{j}^{c}A_{i}^{d}A_{j}^{e}\right].\label{eq:L-int}\end{equation}
In this paper we calculate the three point correlation function at
tree level, thus only the third order interaction Hamiltonian is considered.
For the physical, canonically normalized field it is $\hat{H}_{\mathrm{int}}\equiv\hat{H}_{\mathrm{int}}^{\left(3\right)}+\hat{H}_{\mathrm{int}}^{\left(4\right)}$,
where\begin{eqnarray}
\hat{H}_{\mathrm{int}}^{\left(3\right)} & = & a^{3}\left(\tau\right)\int\mathrm{d}^{3}\mathbf{x}\,\frac{\g}{\sqrt{f}}\f{abc}\partial_{i}\delta\hat{W}_{j}^{a}\delta\hat{W}_{i}^{b}\delta\hat{W}_{j}^{c},\label{eq:H3-def}\\
\hat{H}_{\mathrm{int}}^{\left(4\right)} & = & a^{4}\left(\tau\right)\int\mathrm{d}^{3}\mathbf{x}\,\frac{1}{2}\gf\left(\f{abc}\f{ade}+\f{adc}\f{abe}\right)W_{i}^{b}\delta\hat{W}_{j}^{c}\delta\hat{W}_{i}^{d}\delta\hat{W}_{j}^{e}.\label{eq:H4-def}\end{eqnarray}
The factor $a^{4}$ in these equations is due to $\sqrt{-\mathrm{det}\left[g_{\mu\nu}\right]}$. 

As it is clear from Eqs.~\eqref{eq:H3-def} and \eqref{eq:H4-def}
$g\left(t\right)\equiv\g/\sqrt{f\left(t\right)}$ in Eq.~\eqref{eq:gc-def}
is the strength of self-coupling for the canonically normalized vector
field. To keep quantum calculations under control this coupling must
be ensured to be small. It must also be small for quantum fluctuations
of interacting fields to become classical after horizon exit \citep{Lyth_etal(2006)classicality}.
Furthermore, the perturbation spectrum of massless non-interacting
vector fields is flat if the kinetic function is of the form $f\propto a^{-1\pm3}$
\citep{Dimopoulos_us(2009)_fF2,Dimopoulos_us(2009)_fF2_PRL}. In order
to preserve approximate flatness of the perturbation spectrum, interaction
terms must be small. Because the variation of $f\left(t\right)$ is
exponential at a time $t<t_{\mathrm{s}}$ with $f\left(t_{\mathrm{s}}\right)=1$
by definition, $g=g_{\mathrm{c}}/\sqrt{f}$ is small only when $f\propto a^{-4}$.
We will assume this to be the case for the rest of the paper. 

With the weak self-coupling Eq.~\eqref{eq:U-def} can be expanded
in powers of $\g/\sqrt{f}$. As we are interested in the tree level
contribution to the bispectrum it is enough to keep only the first
order term in Eq.~\eqref{eq:g-def} after which it becomes \citep{Weinberg(2005)nG,Zaldarriaga(2003)nG}\begin{equation}
g_{3}=-i\int_{-\infty}^{0}\mathrm{d}\tau'\left\langle 0\left|\left[\delta\hat{W}_{l}^{f}\left(\mathbf{x}_{1},\tau\right)\delta\hat{W}_{m}^{g}\left(\mathbf{x}_{2},\tau\right)\delta\hat{W}_{n}^{h}\left(\mathbf{x}_{3},\tau\right),\hat{H}_{\mathrm{int}}\left(\tau'\right)\right]\right|0\right\rangle .\label{eq:g3-def}\end{equation}
In what follows $g_{3}^{\left(3\right)}$ will denote the three point
function with the Hamiltonian in Eq.~\eqref{eq:H3-def} and $g_{3}^{\left(4\right)}$
in Eq.~\eqref{eq:H4-def}, so that the total function is $g_{3}=g_{3}^{\left(3\right)}+g_{3}^{\left(4\right)}$.

\subsection{The Three Point Correlation Function from the Quartic Term}

Let us consider first the $g_{3}^{\left(4\right)}$ term. Going to
the momentum space we find\begin{eqnarray}
g_{3}^{\left(4\right)}\left(\mathbf{k}_{1},\mathbf{k}_{2},\mathbf{k}_{3}\right) & = & \left(\f{abc}\f{ade}+\f{adc}\f{abe}\right)W_{i}^{b}\gfi\int_{\tau_{0}}^{\tau_{\mathrm{end}}}\mathrm{d}\tau'a^{8}\left(\tau'\right)\int\frac{\mathrm{d}^{3}q_{1}\mathrm{d}^{3}q_{2}\mathrm{d}^{3}q_{3}}{\left(2\pi\right)^{6}}\delta\left(\mathbf{q}_{1}+\mathbf{q}_{2}+\mathbf{q}_{3}\right)\times\nonumber \\
 &  & \times\mathrm{Re}\left[-i\left\langle 0\left|\delta W_{l}^{f}\left(\mathbf{k}_{1},\tau\right)\delta W_{m}^{g}\left(\mathbf{k}_{2},\tau\right)\delta W_{n}^{h}\left(\mathbf{k}_{3},\tau\right)\delta W_{j}^{c}\left(\mathbf{q}_{1},\tau'\right)\delta W_{i}^{d}\left(\mathbf{q}_{2},\tau'\right)\delta W_{j}^{e}\left(\mathbf{q}_{3},\tau'\right)\right|0\right\rangle \right],\end{eqnarray}
where $\mathrm{Re}\left[\ldots\right]$ denotes the real part and
$\tau_{\mathrm{end}}$ is the conformal time at the end of inflation.
In this expression we used the fact that $W_{i}^{b}$ is slowly rolling
due to the smallness of self coupling term (more on this in section~\ref{sub:spectrum}).
We have also used $f=f_{0}a^{-4}$, where $f_{0}$ is some initial
value. The correlator can be evaluated using Wick's theorem. With
Eq.~\eqref{eq:2pt-corr} after some tedious algebra we find\begin{equation}
g_{3}^{\left(4\right)}\left(\mathbf{k}_{1},\mathbf{k}_{2},\mathbf{k}_{3}\right)=-\left(2\pi\right)^{3}\delta\left(\mathbf{k}_{1}+\mathbf{k}_{2}+\mathbf{k}_{3}\right)\frac{2H^{6}}{\prod_{i}^{3}2k_{i}^{3}}\mathcal{T}_{lmn}^{\left(4\right)fgh}\left(\hat{\mathbf{k}}_{1},\hat{\mathbf{k}}_{2},\hat{\mathbf{k}}_{3}\right)I^{\left(4\right)}\left(k_{1},k_{2},k_{3}\right).\end{equation}
The function $\mathcal{T}_{lmn}^{\left(4\right)fgh}$ depends only
on the direction of three vectors $\left(\mathbf{k}_{1},\mathbf{k}_{2},\mathbf{k}_{3}\right)$
and thus quantifies the anisotropy of the three point correlator function.
The full expression is given by\begin{eqnarray}
\mathcal{T}_{lmn}^{\left(4\right)fgh}\left(\hat{\mathbf{k}}_{1},\hat{\mathbf{k}}_{2},\hat{\mathbf{k}}_{3}\right) & \equiv & W_{m}^{b}\te{lj}1\te{nj}3\left(\f{abh}\f{agf}+\f{agh}\f{abf}\right)+\nonumber \\
 &  & +W_{l}^{b}\te{mj}2\te{nj}3\left(\f{abg}\f{afh}+\f{afg}\f{abh}\right)+\nonumber \\
 &  & +W_{n}^{b}\te{lj}1\te{mj}2\left(\f{abg}\f{ahf}+\f{ahg}\f{abf}\right),\label{eq:T4-def}\end{eqnarray}
where $\te{ij}{}$ is defined in Eq.~\eqref{eq:T-def}. $I^{\left(4\right)}$
is the integral of the form\begin{equation}
I^{\left(4\right)}=\gfi\mathrm{Re}\left[i\int_{\tau_{0}}^{\tau_{\mathrm{end}}}\mathrm{d}\tau'\, a^{8}\left(1-ik_{1}\tau'\right)\left(1-ik_{2}\tau'\right)\left(1-ik_{3}\tau'\right)\mathrm{e}^{ik_{\mathrm{t}}\tau'}\right],\label{eq:I4-def}\end{equation}
with $k_{\mathrm{t}}\equiv k_{1}+k_{2}+k_{3}$. This integral is calculated
in Appendix~\ref{sec:Calculation-of-Intergrals}. Assuming all three
$k$'s crosses the horizon at a similar time it is equal to \begin{eqnarray}
I^{\left(4\right)} & = & \gfi\frac{k_{\mathrm{t}}^{7}H^{-8}}{4!}\left[6\mathrm{e}^{4N_{k}}\left(\frac{1}{3}-K_{1}+K_{2}\right)+2\mathrm{e}^{2N_{k}}\left(K_{1}-3K_{2}-\frac{1}{5}\right)-\right.\nonumber \\
 &  & \left.-\left(\gamma+N_{k}\right)\left(\frac{1}{5}K_{1}-K_{2}-\frac{1}{35}\right)+\frac{1}{300}\left(625K_{2}-137K_{1}+\frac{1019}{49}\right)\right],\label{eq:I4-full}\end{eqnarray}
where\begin{equation}
K_{1}\equiv\frac{\sum_{i>j}^{3}k_{i}k_{j}}{k_{\mathrm{t}}^{2}}\quad\mathrm{and}\quad K_{2}\equiv\frac{\prod_{i}^{3}k_{i}}{k_{\mathrm{t}}^{3}},\label{eq:K1-K2-def}\end{equation}
$\gamma\approx0.577$ is Euler-Mascheroni constant and $N_{k}\equiv-\ln\left(\left|k_{t}\tau_{\mathrm{end}}\right|\right)$
is the number of e-folds from when $k_{\mathrm{t}}$ exits the horizon
to the end of inflation. For the cosmological scales $N_{k}\thicksim60$.
We recognize the dominant term to be proportional to the dominant
term of the three point correlation functions in Refs.~\citep{Zaldarriaga(2003)nG,Seery_etal(2008)nG_Fd_eqn}.
This term is the contribution to the correlation function from the
superhorizon evolution of the fields. The set up in our case is somewhat
different. The authors of Refs.~\citep{Zaldarriaga(2003)nG,Seery_etal(2008)nG_Fd_eqn}
considered a field with constant strength of self-coupling, while
in our case the self-coupling of canonically normalized fields is
varying with time, $g=g_{\mathrm{c}}/\sqrt{f}\propto a^{2}$. From
Eq.~\eqref{eq:I4-full} we see that this variation enhances additional
modes, both, when they are created at the horizon exit and during
the evolution of the field outside the horizon. However, these modes
are subdominant. Taking only the dominant contribution to $g_{3}^{\left(4\right)}$
we find\begin{equation}
g_{3}^{\left(4\right)}=-\left(2\pi\right)^{3}\delta\left(\mathbf{k}_{1}+\mathbf{k}_{2}+\mathbf{k}_{3}\right)\mathcal{T}_{lmn}^{\left(4\right)fgh}\left(\hat{\mathbf{k}}_{1},\hat{\mathbf{k}}_{2},\hat{\mathbf{k}}_{3}\right)\frac{\sum_{i}^{3}k_{i}^{3}}{\prod_{i}k_{i}^{3}}\,\frac{\g^{2}}{f_{\mathrm{end}}}\frac{H^{2}}{48}.\label{eq:g3-4-final}\end{equation}
To evaluate this equation we used $f_{\mathrm{end}}=f_{k}\exp\left(-4N_{k}\right)$,
where $f_{k}$ and $f_{\mathrm{end}}$ are the values of $f$ at the
horizon crossing and the end of inflation respectively.

\subsection{The Three Point Correlation Function from the Cubic Term}

From Eq.~\eqref{eq:H3-def} we expect that the three point correlation
function from the cubic term $g_{3}^{\left(3\right)}$ is suppressed
by a factor of $p$ compared to $g_{3}^{\left(4\right)}$, where $p=k/a$
is the modulus of the physical momentum. In this section we show that
this is indeed the case. $g_{3}^{\left(3\right)}$ is calculated along
the same lines as $g_{3}^{\left(4\right)}$. Taking a Fourier transform
of Eq.~\eqref{eq:H3-def} from Eq.~\eqref{eq:g3-def} we find\begin{eqnarray}
g_{3}^{\left(3\right)} & = & \frac{\g}{\sqrt{f_{0}}}\f{abc}\int_{\tau_{0}}^{\tau_{\mathrm{end}}}\mathrm{d}\tau'\, a^{5}\left(\tau'\right)\int\frac{\mathrm{d}^{3}q_{1}\mathrm{d}^{3}q_{2}\mathrm{d}^{3}q_{3}}{\left(2\pi\right)^{6}}\delta\left(\mathbf{q}_{1}+\mathbf{q}_{2}+\mathbf{q}_{3}\right)q_{1i}\times\nonumber \\
 &  & \times2i\,\mathrm{Im}\left[\left\langle 0\left|\delta W_{l}^{f}\left(\mathbf{k}_{1},\tau\right)\delta W_{m}^{g}\left(\mathbf{k}_{2},\tau\right)\delta W_{n}^{h}\left(\mathbf{k}_{3},\tau\right)\delta W_{j}^{a}\left(\mathbf{q}_{1},\tau'\right)\delta W_{i}^{b}\left(\mathbf{q}_{2},\tau'\right)\delta W_{j}^{c}\left(\mathbf{q}_{3},\tau'\right)\right|0\right\rangle \right],\end{eqnarray}
where $\mathrm{Im}\left[\ldots\right]$ denotes the imaginary part.
Using Wick's theorem and Eq.~\eqref{eq:2pt-corr} we calculate\begin{equation}
g_{3}^{\left(3\right)}=-\left(2\pi\right)^{3}\delta\left(\mathbf{k}_{1}+\mathbf{k}_{2}+\mathbf{k}_{3}\right)\frac{2H^{6}}{\prod_{i}^{3}2k_{i}^{3}}\mathcal{T}_{lmn}^{\left(3\right)fgh}\left(\mathbf{k}_{1},\mathbf{k}_{2},\mathbf{k}_{3}\right)I^{\left(3\right)}\left(k_{1},k_{2},k_{3}\right),\end{equation}
where the anisotropy of the three point correlation function is given
by\begin{eqnarray}
\mathcal{T}_{lmn}^{\left(3\right)fgh}\left(\mathbf{k}_{1},\mathbf{k}_{2},\mathbf{k}_{3}\right) & = & \f{fgh}\left[\te{lj}1\te{mi}2\te{jn}3\left(\mathbf{p}_{1}-\mathbf{p}_{3}\right)_{i}+\right.\nonumber \\
 &  & \qquad+\te{lj}1\te{jm}2\te{ni}3\left(\mathbf{p}_{2}-\mathbf{p}_{1}\right)_{i}+\nonumber \\
 &  & \qquad\left.+\te{li}1\te{mj}2\te{jn}3\left(\mathbf{p}_{3}-\mathbf{p}_{2}\right)_{i}\right].\label{eq:T3-def}\end{eqnarray}
$\mathbf{p}\equiv\mathbf{k}/a_{\mathrm{end}}$ in this expression
is the physical momentum, $a_{\mathrm{end}}$ is the scale factor
at the end of inflation and\begin{equation}
I^{\left(3\right)}\equiv-ia_{\mathrm{end}}\frac{\g}{\sqrt{f_{0}}}\mathrm{Im}\left[\int_{\tau_{0}}^{\tau_{\mathrm{end}}}\mathrm{d}\tau'\, a^{5}\left(\tau'\right)\left(1-ik_{1}\tau'\right)\left(1-ik_{2}\tau'\right)\left(1-ik_{3}\tau'\right)\mathrm{e}^{ik_{\mathrm{t}}\tau'}\right].\label{eq:I3-def}\end{equation}
Using the method explained in Appendix~\ref{sec:Calculation-of-Intergrals}
it is calculated to be\begin{equation}
I^{\left(3\right)}=-ia_{\mathrm{end}}\frac{\g}{\sqrt{f_{0}}}k_{\mathrm{t}}^{4}H^{-5}\left[\mathrm{e}^{N_{k}}\left(\frac{1}{3}-K_{1}+K_{2}\right)-\frac{\pi}{4}\left(\frac{1}{4}-K_{1}+2K_{2}\right)\right].\end{equation}
The first, dominant term, is due to the evolution after horizon exit.
Neglecting the subdominant term $g_{3}^{\left(3\right)}$ becomes\begin{equation}
g_{3}^{\left(3\right)}=i\left(2\pi\right)^{3}\delta\left(\mathbf{k}_{1}+\mathbf{k}_{2}+\mathbf{k}_{3}\right)\mathcal{T}_{lmn}^{\left(3\right)fgh}\left(\mathbf{k}_{1},\mathbf{k}_{2},\mathbf{k}_{3}\right)\frac{\sum_{i}^{3}k_{i}^{3}}{\prod_{i}k_{i}^{3}}\,\frac{\g}{\sqrt{f_{\mathrm{end}}}}\frac{H^{2}}{12}.\label{eq:g3-3-final}\end{equation}
The anisotropic term $\mathcal{T}^{\left(4\right)}$ from quartic
interactions in Eq.~\eqref{eq:T4-def} is proportional to the homogeneous
part of the vector field $W>H$,%
\footnote{\label{fn:W-g-H-bound}For the perturbative approach to be valid $W>\delta W$
must hold. The typical value of the field perturbation is $\delta W\sim\sqrt{\mathcal{P}_{+}}=H/2\pi$,
resulting in $W>H$.%
} while $\mathcal{T}^{\left(3\right)}$ from cubic interactions is
proportional to the physical momentum $p$, which for cosmological
scales are $p\ll H$. Thus $\left|g_{3}^{\left(4\right)}\right|\gg\left|g_{3}^{\left(3\right)}\right|$
and the dominant contribution to the three point correlation function
is from quartic terms. Moreover, the dominant contribution in $g^{\left(4\right)}$
itself is from the classical evolution of fields.

\section{Bispectrum from the Classical Evolution\label{sec:classical}}

\subsection{Classicality}

In the last section we used perturbative quantum field theory to calculate
correlators of the field perturbation. As the results in Eqs.~\eqref{eq:g3-4-final}
and \eqref{eq:g3-3-final} show those correlators are dominated by
the interaction of fields after horizon exit. Furthermore, as correlators
with derivative couplings are suppressed by the factor $k/a\ll H$,
it suggests that we can obtain the dominant contribution to correlators
by a simpler method: from the classical equation of motion of the
homogeneous field.

It is well known that after a mode $\mathbf{k}$ of a free light quantum
field crosses the horizon, i.e. when $k/aH\rightarrow0$, the phase
of the mode function becomes constant and field operator in the Heisenberg
picture can be written in a form $\chi_{k}\left(\tau\right)\left(\hat{a}_{\mathbf{k}}+\hat{a}_{-\mathbf{k}}^{\dagger}\right)$,
where $\chi_{k}$ is made real by an arbitrary phase rotation \citep{Starobinsky(1982)Classicality,Polarski_Starobinski(1995)}.
In this limit all commutators of fields vanish and the eigenvector
of a field operator at some particular time remains an eigenvector
thereafter. This is a cosmological analogue of quantum decoherence.
When this happens, quantum fields are well described by classical
stochastic functions and we say that the field enters into `classical
evolution'. This is true for light free quantum fields. Canonical
massless vector fields, however, do not become classical as their
Lagrangian is invariant under conformal transformation to flat space-time.
This is a case, for example, with $U\left(1\right)$ vector field
with minimal kinetic term \citep{TurnerWidrow1988}. In our case,
conformal invariance of the vector field is broken by the time varying
kinetic function $f$ in Eq.~\eqref{eq:Lagrangian}. But the question
remains whether self-interaction terms do not prevent the non-Abelian
field from becoming classical. In Ref.~\citep{Lyth_etal(2006)classicality}
it was shown that after horizon crossing the interacting field does
become classical if the interaction is weak; specifically if $\hat{U}$
in Eq.~\eqref{eq:U-def} is sufficiently close to unity, $\hat{U}\simeq\hat{1}$.
As was discussed before this can only happen if $f$ is a decreasing
function in time. With a constraint of the flat perturbation spectrum
this means $f\propto a^{-4}$.

In this section we show that correlator functions of the non-Abelian
vector field perturbation can be calculated using the classical equation
of motion.

\subsection{The Equation of Motion and the Power Spectrum\label{sub:spectrum}}

Extremising the action with the Lagrangian in Eq.~\eqref{eq:Lagrangian}
we obtain the field equation for non-Abelian vector fields\begin{equation}
\left[\partial_{\lambda}+\partial_{\lambda}\ln\sqrt{-\det\left[g_{\mu\nu}\right]}\right]\left(fF_{h}^{\lambda\kappa}\right)-f\g\f{abh}A_{\rho}^{b}F_{a}^{\rho\kappa}=0,\label{eq:Fd-eqn}\end{equation}
Taking the spatial component of this equation $\kappa=i$ and adopting
a temporal gauge $A_{0}^{a}=0$ we get\begin{equation}
\ddot{A}_{i}^{h}+\left(H+\frac{\dot{f}}{f}\right)\dot{A}_{i}^{h}-a^{-2}\left(\partial_{j}\partial_{j}A_{i}^{h}-\partial_{i}\partial_{j}A_{j}^{h}\right)-a^{-2}\g\f{abh}\left[2\left(A_{j}^{a}\partial_{j}A_{i}^{a}+\partial_{j}A_{j}^{a}A_{i}^{b}\right)-\g\f{ade}A_{j}^{b}A_{j}^{d}A_{i}^{e}\right]=0.\label{eq:EoM-Ai-gen}\end{equation}
We are interested in superhorizon evolution of the vector field perturbation.
On these scales derivative terms are negligible and Eq.~\eqref{eq:EoM-Ai-gen}
can be written as\begin{equation}
\ddot{A}_{i}^{h}+\left(H+\frac{\dot{f}}{f}\right)\dot{A}_{i}^{h}+\g^{2}\f{abh}\f{ade}a^{-2}A_{j}^{b}A_{j}^{d}A_{i}^{e}=0,\end{equation}
which is the same as that of the homogeneous mode of the field. Changing
to the physical canonically normalized vector field in Eq.~\eqref{eq:W-def},
with $f\propto a^{-4}$, this equation is transformed into\begin{equation}
\ddot{W}_{i}^{h}+3H\dot{W}_{i}^{h}+\gf\f{abh}\f{ade}W_{j}^{b}W_{j}^{d}W_{i}^{e}=0,\label{eq:W-EoM-hom}\end{equation}
which is reminiscent of an interacting scalar field. As fields in
Eq.~\eqref{eq:W-EoM-hom} are intended to generate the curvature
perturbation, they must retain an approximate scale invariance in
accordance with observations. For this to be the case $W_{i}^{h}$
must be almost a free field, in other words $W_{i}^{h}$ must be rolling
slowly. In analogy to Ref.~\citep{Seery_etal(2005)multiFd} we introduce
slow-roll parameters\begin{equation}
\epsilon_{ij}^{ab}\equiv\frac{\dot{W}_{i}^{a}\dot{W}_{j}^{b}}{2\mpl H}\quad\mathrm{and}\quad\eta_{ij}^{ab}\equiv\frac{V_{ij}^{ab}}{3H^{2}},\label{eq:slow-roll-def}\end{equation}
where $V_{ij}^{ab}\equiv\partial V_{i}^{a}/\partial W_{j}^{b}$ and\begin{equation}
V_{i}^{h}\equiv\gf\f{abh}\f{ade}W_{j}^{b}W_{j}^{d}W_{i}^{e}\propto g^{2}\left(t\right)\end{equation}
and require $\left|\epsilon_{ij}^{ab}\right|<1$ and $\left|\eta_{ij}^{ab}\right|\sim\frac{g_{\mathrm{c}}^{2}}{3f}\left(\frac{W}{H}\right)^{2}<1$.
As structure constants are of order unity, the slow-roll conditions
mean that the strength of self-coupling $g\left(t\right)\equiv\g/\sqrt{f\left(t\right)}$
in Eq.~\eqref{eq:gc-def} of canonically normalized field is small,
i.e. $\fgh\ll1$. This is easily achieved when cosmological scales
exit the horizon. Because $f\propto a^{-4}$ is an exponentially decaying
function with $f\left(t_{\mathrm{s}}\right)=1$, the self-coupling
$\fgh$ is exponentially suppressed. Note however that although $\left|\eta_{ij}^{ab}\right|\ll1$
is easily satisfied when cosmological scales exit the horizon, this
condition must hold up until $t_{\mathrm{s}}$. Even if the evolution
becomes strongly non-linear after cosmological scales crosses the
horizon, all scales are affected. Thus not only $\fgh$ must be small
at horizon crossing but it must remain small when $f\left(t_{\mathrm{s}}\right)=1$,
i.e. $\g^{2}<1$. We assume this to be the case.

Following Ref.~\citep{Seery_etal(2008)nG_Fd_eqn} we decompose the
vector field as \begin{equation}
W^{a}=W_{0}^{a}+\delta W_{1}^{a}+\frac{1}{2}\delta W_{2}^{a}+\ldots,\label{eq:W-expansion}\end{equation}
where the field notation without space indices means the modulus,
e.g. $W^{a}\equiv\left|W_{i}^{a}\right|$. The first term in this
expression is the homogeneous field. For the rest of the paper we
will have no use of the total vector field $W_{i}^{a}$, thus we drop
out the subscript `$0$' from the homogeneous mode and denote it simply
by $W_{i}^{a}$. The second term in Eq.~\eqref{eq:W-expansion} is
the perturbation and later terms are higher orders in $\delta W_{1}^{a}$.
This expansion is not unique and to determine $\delta W_{1}^{a}$
some auxiliary conditions need to be imposed \citep{Malik(2003)2ndOrder,Seery_etal(2008)nG_Fd_eqn}.
We choose $\delta W_{1}^{a}$ in such a way that its equation of motion
is linear, i.e. Eq.~\eqref{eq:W-EoM-hom} without the last term.
It follows then that $\delta W_{1}^{a}$ obeys the Gaussian statistics
and its two point correlation function is \begin{equation}
\left\langle \delta W_{1i}^{a}\left(\mathbf{k}\right)\delta W_{1j}^{b}\left(\mathbf{k}'\right)\right\rangle =\left(2\pi\right)^{3}\delta\left(\mathbf{k}+\mathbf{k}'\right)\frac{2\pi^{2}}{k^{3}}\delta_{ab}\te{ij}{}\mathcal{P}_{+}^{a}\left(k\right),\label{eq:W1-2pnt-def}\end{equation}
where $\te{ij}{}$ is defined in Eq.~\eqref{eq:T-def} and the Fourier
transform of $\delta W_{1i}^{a}$ is defined in Eq.~\eqref{eq:W-Fourier}.
We also used the fact that there is no correlation between left- and
right-handed modes, hence the Kronecker delta $\delta_{ab}$. In de
Sitter inflation the power spectrum is scale invariant $\mathcal{P}_{+}^{a}=\left(H/2\pi\right)^{2}$.
In the slow-roll inflation the spectrum acquires weak scale dependence
due to slowly increasing horizon size. Assuming approximately constant
$\dot{H}/H^{2}$ it is\begin{equation}
\mathcal{P}_{+}^{a}=\left(\frac{H}{2\pi}\right)^{2}\left(\frac{k}{aH}\right)^{-2\epsilon}.\label{eq:W-spectrum0}\end{equation}
The spectral tilt `$-2\epsilon$' is due to the slight increase of
the Hubble horizon during slow-roll inflation, which is parametrized
by $\epsilon\equiv-\dot{H}/H^{2}$. Because $\epsilon>0$, each subsequent
$k$ mode crosses a horizon of larger size making the amplitude of
perturbation smaller.

The spectrum in Eq.~\eqref{eq:W-spectrum0} is for the non-interacting
part $\delta W_{1}$ of the field perturbation. However, the total
power spectrum will have additional scale dependence

\begin{equation}
\mathcal{P}_{+}^{a}=\left(\frac{H}{2\pi}\right)^{2}\left(\frac{k}{aH}\right)^{-2\epsilon+2\left|\eta\right|},\label{eq:W1-spectrum}\end{equation}
where $\left|\eta\right|=\left|\eta_{ij}^{ab}\right|$ is the modulus
of the slow-roll parameter matrix in Eq.~\eqref{eq:slow-roll-def}.
The second term in the exponent of the scale dependent factor on the
right-hand-side of the above equation is caused by interactions in
the last term of Eq.~\eqref{eq:W-EoM-hom}. Due to interactions each
$k$ mode is not frozen after horizon exit but evolves slowly. As
larger modes spend less time outside the horizon they are less affected,
which introduces additional $k$ dependence. Both $\epsilon$ and
$\left|\eta\right|$ in Eq.~\eqref{eq:W1-spectrum} are evaluated
at horizon crossing. Although $\left|\eta\right|\propto f^{-1}$ is
a function of time this does not introduce additional scale dependence
as all modes after horizon crossing are affected the same way by the
evolution of $\left|\eta\right|$. However, $\left|\eta\right|\propto f^{-1}$
means that for cosmological scales $\left|\eta\right|$ is exponentially
suppressed and the $\epsilon$ term dominates the spectral tilt. In
principle $\left|\eta\right|$ introduces an anisotropic scale dependence.
But as this term is subdominant, the direction dependence of $\left|\eta\right|$
is suppressed.

The presence of non-linear term in Eq.~\eqref{eq:W-EoM-hom} makes
the vector field perturbation non-Gaussian. As we have chosen $\delta W_{1}$
to satisfy Gaussian statistics, non-Gaussianity is encapsulated in
$\delta W_{2}$. Thus $\delta W_{2}$ satisfies the full non-linear
equation. However, the curvature perturbation in the Universe is predominantly
Gaussian. So if vector fields are to generate the dominant contribution
to the curvature perturbation, they must be predominantly Gaussian
too, that is $\delta W_{2}^{a}<\delta W_{1}^{a}$. With this condition
$\delta W_{2i}^{a}$ can be seen as the second order perturbation.
Thus perturbing Eq.~\eqref{eq:W-EoM-hom} to the second order we
find the equation of motion for $\delta W_{2i}^{a}$

\begin{equation}
\delta\ddot{W}_{2i}^{h}+3H\delta\dot{W}_{2i}^{h}+2V_{2i}^{h}=0.\label{eq:W2-EoM}\end{equation}
Dropping out terms proportional to the slow-roll parameters in Eq.~\eqref{eq:slow-roll-def}
we write\begin{equation}
V_{2i}^{h}\equiv\frac{\partial^{2}V_{i}^{h}}{\partial W_{m}^{a}\partial W_{n}^{b}}\delta W_{1m}^{a}\delta W_{1n}^{b}=\gf\f{abh}\f{adc}\left[W_{i}^{c}\delta W_{1j}^{b}\delta W_{1j}^{d}+W_{j}^{d}\delta W_{1j}^{b}\delta W_{1i}^{c}+W_{j}^{b}\delta W_{1j}^{d}\delta W_{1i}^{c}\right]\label{eq:V2-def}\end{equation}
Assuming slow-roll holds we may also drop the first term in Eq.~\eqref{eq:W2-EoM}
and write\begin{equation}
\delta\dot{W}_{2i}^{h}=-\frac{2V_{2i}^{h}}{3H}.\end{equation}
Taking the Fourier transform it becomes\begin{equation}
\delta\dot{W}_{2i}^{h}\left(\mathbf{k}\right)=-\frac{2}{3H}\int\frac{\mathrm{d}^{3}q_{1}\mathrm{d}^{3}q_{2}}{\left(2\pi\right)^{3}}\delta\left(\mathbf{k}-\mathbf{q}_{1}-\mathbf{q}_{2}\right)V_{2i}^{h}\left(\mathbf{q}_{1},\mathbf{q}_{2}\right),\label{eq:dW-eqn}\end{equation}
where $V_{2i}^{h}\left(\mathbf{q}_{1},\mathbf{q}_{2}\right)$ is\begin{equation}
V_{2i}^{h}\left(\mathbf{q}_{1},\mathbf{q}_{2},t\right)=\frac{g_{\mathrm{c}}^{2}}{f\left(t\right)}\f{abh}\f{adc}\left[W_{i}^{c}\delta W_{1j}^{b}\left(\mathbf{q}_{1}\right)\delta W_{1j}^{d}\left(\mathbf{q}_{2}\right)+W_{j}^{d}\delta W_{1j}^{b}\left(\mathbf{q}_{1}\right)\delta W_{1i}^{c}\left(\mathbf{q}_{2}\right)+W_{j}^{b}\delta W_{1j}^{d}\left(\mathbf{q}_{1}\right)\delta W_{1i}^{c}\left(\mathbf{q}_{2}\right)\right].\label{eq:V2-Fourier}\end{equation}
As we are interested in the superhorizon evolution of the field perturbation,
to find $\delta W_{2i}^{h}$ we integrate Eq.~\eqref{eq:dW-eqn}
from the horizon exit at $t_{k}$, where $k/a\left(t_{k}\right)H=1$,
to some later time $t$. Because $W_{i}^{a}$ is slowly rolling and
$\delta W_{1i}^{a}$ is constant by definition with $H\approx\mathrm{const}$,
the only time dependent term in Eq.~\eqref{eq:V2-Fourier} is $f\propto a^{-4}$.
Thus solving Eq.~\eqref{eq:dW-eqn} we find\begin{equation}
\delta W_{2i}^{h}\left(\mathbf{k},t\right)=-\frac{1}{6H^{2}}\int\frac{\mathrm{d^{3}q_{1}}\mathrm{d}^{3}q_{2}}{\left(2\pi\right)^{3}}\delta\left(\mathbf{k}-\mathbf{q}_{1}-\mathbf{q}_{2}\right)\te{ij}{}V_{2j}^{h}\left(\mathbf{q}_{1},\mathbf{q}_{2},t\right).\label{eq:W2-solution}\end{equation}

From this solution we can find the bound on the strength of self-coupling
$\fgh$ for the condition $\delta W_{2}<\delta W_{1}$ to be consistent.
Putting $\delta W_{1}\sim H$ into Eqs.~\eqref{eq:V2-Fourier} and
\eqref{eq:W2-solution} it follows $\frac{\g^{2}}{f}W<H$. This ensures
consistency of using second order perturbation theory to calculate
$\delta W_{2}$ and that perturbations of vector fields are predominantly
Gaussian.

\subsection{The Three-Point Correlation Function}

The three point correlation function from the classical evolution
of the field is\begin{equation}
\gcl\left(\mathbf{k}_{1},\mathbf{k}_{2},\mathbf{k}_{3}\right)\equiv\left\langle \delta W_{l}^{f}\left(\mathbf{k}_{1}\right)\delta W_{m}^{g}\left(\mathbf{k}_{2}\right)\delta W_{n}^{h}\left(\mathbf{k}_{3}\right)\right\rangle .\label{eq:gcl-def}\end{equation}
Because $\delta W_{1}\gg\delta W_{2}$ the largest contribution to
$\gcl$ comes from the term of the form $\left\langle \delta W_{1}\delta W_{1}\frac{1}{2}\delta W_{2}\right\rangle \propto\frac{1}{2}\left\langle \delta W_{1}\delta W_{1}\delta W_{1}\star\delta W_{1}\right\rangle $,
where a star denotes convolution. As one has to be careful in keeping
track of indices we write the dominant term of $\gcl$ explicitly\begin{equation}
\gcl=\frac{1}{2}\left[\left\langle \w{1l}f\left(\mathbf{k}_{1}\right)\w{1m}g\left(\mathbf{k}_{2}\right)\w{2n}h\left(\mathbf{k}_{3}\right)\right\rangle +\left\langle \w{1l}f\left(\mathbf{k}_{1}\right)\w{2m}g\left(\mathbf{k}_{2}\right)\w{1n}h\left(\mathbf{k}_{3}\right)\right\rangle +\left\langle \w{2l}f\left(\mathbf{k}_{1}\right)\w{1m}g\left(\mathbf{k}_{2}\right)\w{1n}h\left(\mathbf{k}_{3}\right)\right\rangle \right].\end{equation}

To evaluate this expression we use Eq.~\eqref{eq:W2-solution} and
Wick's theorem to express four point functions in terms of products
of two point ones. After tedious algebra and using Eqs.~\eqref{eq:W1-2pnt-def}
we obtain\begin{eqnarray}
\gcl & = & -\left(2\pi\right)^{3}\delta\left(\mathbf{k}_{1}+\mathbf{k}_{2}+\mathbf{k}_{3}\right)\mathcal{T}_{lmn}^{\left(4\right)fgh}\left(\hat{\mathbf{k}}_{1},\hat{\mathbf{k}}_{2},\hat{\mathbf{k}}_{3}\right)\frac{\sum_{i}^{3}k_{i}^{3}}{\prod_{i}k_{i}^{3}}\,\frac{\g^{2}}{f_{\mathrm{end}}}\frac{4\pi^{4}\mathcal{P}_{+}^{2}}{12H^{2}},\label{eq:g3-cl-final}\end{eqnarray}
where the anisotropy tensor $\mathcal{T}_{lmn}^{\left(4\right)fgh}$
is defined in Eq.~\eqref{eq:T4-def} and we dropped the gauge index
from $\mathcal{P}_{+}^{a}$ as all vector fields have the same spectrum.
Taking $\mathcal{P}_{+}=\left(H/2\pi\right)^{2}$ we recover exactly
the same result as obtained by a more tedious calculation in the quantum
in-in formalism with the dominant term in Eq.~\eqref{eq:g3-4-final}.

\section{The Curvature Perturbation\label{sec:curvature-perturbation}}

\subsection{$\delta N$ Formula}

In the above we have calculated correlators of the field perturbation.
However, field perturbation is not an observable, but the metric perturbation
is. We choose a uniform density slicing in which the perturbation
of the metric on superhorizon scales is described by the intrinsic
curvature $\zeta$. The easiest way to calculate $\zeta$ is using
the $\delta N$ formula \citep{Starobinsky(1986)dN,Sasaki_Stewart(1995)dN}.
This formula was first extended to include vector fields in Ref.~\citep{Dimopoulos_etal_anisotropy(2008)}
and used for non-Abelian fields in Refs.~\citep{Bartolo_etal(2009)_Bispectrum,Bartolo_etal(2009)Trispectrum}\begin{equation}
\zeta\left(\mathbf{x},t\right)=N_{\phi}\delta\phi+N_{i}^{a}\delta W_{i}^{a}+\frac{1}{2}N_{\phi\phi}\left(\delta\phi\right)^{2}+N_{\phi i}^{a}\delta\phi\delta W_{i}^{a}+\frac{1}{2}N_{ij}^{ab}\delta W_{i}^{a}\delta W_{j}^{b}+\ldots,\label{eq:zeta-def}\end{equation}
where\begin{equation}
N_{\phi}\equiv\frac{\partial N}{\partial\phi},\; N_{i}^{a}\equiv\frac{\partial N}{\partial W_{i}^{a}},\; N_{\phi i}^{a}\equiv\frac{\partial^{2}N}{\partial\phi\partial W_{i}^{a}}\;\mathrm{and}\; N_{ij}^{ab}\equiv\frac{\partial^{2}N}{\partial W_{i}^{a}\partial W_{j}^{b}}.\label{eq:dNs-defs}\end{equation}
$N$ in these expressions is the number of e-foldings of local expansion
from the initial flat hypersurface to the final uniform density hypersurface
at final time $t$ when $\zeta$ becomes constant. In Refs.~\citep{Bartolo_etal(2009)_Bispectrum,Bartolo_etal(2009)Trispectrum}
$t$ was taken to be just after the horizon crossing. However, as
Eq.~\eqref{eq:g3-4-final} shows, the bispectrum of the field perturbation
is actually dominated by the interaction of classical fields during
classical evolution outside the horizon.

Derivatives in Eq.~\eqref{eq:dNs-defs} are taken with respect to
homogeneous fields. The precise form of these derivatives depends
on the mechanism through which the field perturbation generates $\zeta$.
$\delta\phi$ in this equation is the perturbation of some scalar
field if any of such fields contribute to the curvature perturbation.
In this paper, by keeping only the second and fifth terms of the right-hand-side
of Eq.~\eqref{eq:zeta-def} we assume that predominantly vector fields
contribute to $\zeta$ and any other source is negligible.

\subsection{The Spectrum\label{sub:zeta-Spectrum}}

Let us first consider the two point correlation function of $\zeta$.
In Fourier space we may write\begin{equation}
\left\langle \zeta\left(\mathbf{k}\right)\zeta\left(\mathbf{k}'\right)\right\rangle =N_{i}^{a}N_{j}^{b}\left\langle \delta W_{i}^{a}\left(\mathbf{k}\right)\delta W_{j}^{b}\left(\mathbf{k}'\right)\right\rangle \equiv\left(2\pi\right)^{2}\delta\left(\mathbf{k}+\mathbf{k}'\right)\frac{2\pi^{2}}{k^{3}}\mathcal{P}_{\zeta}\left(\mathbf{k}\right),\end{equation}
where $\delta W_{i}^{a}=\delta W_{1i}^{a}+\frac{1}{2}\delta W_{2i}^{a}+\dots$.
From Eq.~\eqref{eq:W1-2pnt-def} we find\begin{equation}
\mathcal{P}_{\zeta}\left(\mathbf{k}\right)=\sum_{a,b}\delta_{ab}N_{i}^{a}N_{j}^{b}\te{ij}{}\mathcal{P}_{+}^{a}\left(k\right).\label{eq:Pz-1}\end{equation}
Due to the presence of $\te{ij}{}$ in this expression the power spectrum
of $\zeta$ has an angular modulation. The isotropic part is \begin{equation}
\mathcal{P}_{\zeta}^{\mathrm{iso}}\equiv\mathcal{P}_{+}\sum_{a}N_{a}^{2},\label{eq:Pz-iso-def}\end{equation}
where the sum is over all gauge fields and $N_{a}$ is the absolute
value $N_{a}\equiv\left|N_{i}^{a}\right|$. We also used the fact
that power spectra $\mathcal{P}_{+}^{a}=\mathcal{P}_{+}$ are the
same for all fields. Then the total spectrum of $\zeta$ is\begin{equation}
\mathcal{P}_{\zeta}\left(\mathbf{k}\right)=\mathcal{P}_{\zeta}^{\mathrm{iso}}\left[1-\frac{\sum_{a}\left(\mathbf{N}^{a}\cdot\hat{\mathbf{k}}\right)^{2}}{\sum_{a}N_{a}^{2}}\right].\label{eq:Pz-nonAb}\end{equation}
If $a=1$ this expression reduces to Eq.~\eqref{eq:Pz-aniso-def}
with quadrupole anisotropy of an amplitude $g_{\zeta}=-1$ and such
a large anisotropy is ruled out by observations. Thus, a massless
$U\left(1\right)$ vector field cannot generate the total $\zeta$.
However, with the large number of randomly oriented vector fields,
the anisotropy is suppressed by the number of fields. To see this,
note that $N$ is proportional to vector fields $W^{a}$. Since all
vector fields satisfy the same equation of motion, assuming similar
initial conditions it is reasonable to expect that contributions of
all $W^{a}$ to $N$ are of the same order. In this case $N_{a}$'s
are of the same order too, in particular $N_{W}\equiv N_{a}\sim N_{b}$
for all $a$ and $b$. Then $\sum_{a}N_{a}^{2}=\mathcal{N}N_{W}^{2}$,
where $\mathcal{N}$ is the number of fields and Eq.~\eqref{eq:Pz-nonAb}
becomes\begin{equation}
\mathcal{P}_{\zeta}\left(\mathbf{k}\right)=\mathcal{N}\mathcal{P}_{+}N_{W}^{2}\left[1-\frac{1}{\mathcal{N}}\sum_{a}^{\mathcal{N}}\left(\hat{\mathbf{W}}^{a}\cdot\hat{\mathbf{k}}\right)^{2}\right],\label{eq:Pz-gen}\end{equation}
where $\hat{\mathbf{W}}^{a}$ are unit vectors along the directions
of homogeneous vector fields $W_{i}^{a}$ and we used the fact that
$\hat{\mathbf{W}}^{a}=\hat{\mathbf{N}}^{a}\equiv\mathbf{N}^{a}/N_{a}$.
With $\mathcal{N}$ vector fields, the anisotropic part of the spectrum
is a sum of $\mathcal{N}$ quadrupoles. If these are randomly oriented,
the anisotropy is suppressed by $\mathcal{N}^{-1}$. Thus, with the
large enough non-Abelian symmetry group the total curvature perturbation
can be generated solely by vector fields. In view of contradicting
conclusions of Refs.~\citep{Groeneboom_etal(2009)anisotropy2,Hanson_etal(2010)statAnis}
we consider two bounds on $g_{\zeta}$ to estimate $\mathcal{N}$.
If we accept that a systematic effect causing the large anisotropy
in the spectrum ($0.29\pm0.031$ as claimed in Ref.~\citep{Groeneboom_etal(2009)anisotropy2})
is unknown, the bound on the anisotropy in $\mathcal{P}_{\zeta}$
of the primordial origin can be taken to be $g_{\zeta}<0.29$. In
such a case four vector fields $\mathcal{N}=4$ is enough for this
bound to be satisfied. If, on the other hand, the large observed anisotropy
in $\mathcal{P}_{\zeta}$ is caused by the WMAP beam asymmetry, as
claimed in Ref.~\citep{Hanson_etal(2010)statAnis}, then the corrected
bound on primordial statistical anisotropy gives $\left|g_{\zeta}\right|<0.07$
\citep{Ma_etal(2011)statAnis}. To satisfy this bound $\mathcal{N}\geq15$
is needed. These estimates for $\mathcal{N}$ are made assuming random
orientation of the homogeneous vector fields and their similar magnitudes.
If, however, all vector fields are parallel or one of the $N_{a}$'s
is dominant, then $g_{\zeta}=-1$ and such configuration is excluded.

It is also possible to generate statistically isotropic curvature
perturbation by considering a triad of orthogonal vector fields with
equal norm. This configuration in a context of vector inflation was
studied in Refs.~\citep{Maleknejad_etal(2011)short,Maleknejad_etal(2011)long},
where $SU\left(2\right)$ group is considered. However, we feel that
a scenario with random orientation of larger number of fields is a
more natural setup. In addition, such setup also has an advantage
of providing observational signature, that is non-negligible statistical
anisotropy.

\subsection{The Bispectrum}

In this section we calculate the bispectrum $B_{\zeta}$ at the end
of inflation. In momentum space it is defined by\begin{equation}
\left\langle \zeta\left(\mathbf{k}_{1}\right)\zeta\left(\mathbf{k}_{2}\right)\zeta\left(\mathbf{k}_{3}\right)\right\rangle =\left(2\pi\right)^{3}\delta\left(\mathbf{k}_{1}+\mathbf{k}_{2}+\mathbf{k}_{3}\right)B_{\zeta}\left(\mathbf{k}_{1},\mathbf{k}_{2},\mathbf{k}_{3}\right).\end{equation}
Non-vanishing $B_{\zeta}$ is the result of two contributions. The
first contribution, let us denote it by $B_{\zeta1}$, is from the
non-Gaussian field perturbations due to self-interactions of the vector
fields. The second, $B_{\zeta2}$, is due to the non-linear terms
in the $\delta N$ formula in Eq.~\eqref{eq:zeta-def}.

Let us start by calculating the first contribution. From Eq.~\eqref{eq:zeta-def}

\begin{equation}
\left\langle \zeta\left(\mathbf{k}_{1}\right)\zeta\left(\mathbf{k}_{2}\right)\zeta\left(\mathbf{k}_{3}\right)\right\rangle \supset N_{l}^{f}N_{m}^{g}N_{n}^{h}\left\langle \delta W_{l}^{f}\left(\mathbf{k}_{1}\right)\delta W_{m}^{g}\left(\mathbf{k}_{2}\right)\delta W_{n}^{h}\left(\mathbf{k}_{3}\right)\right\rangle .\end{equation}
The three point correlation function of the field perturbation was
calculated in sections \ref{sec:quantum} and \ref{sec:classical}.
Taking the result in Eq.~\eqref{eq:g3-cl-final} the bispectrum $B_{\zeta1}$
becomes\begin{equation}
B_{\zeta1}=-g_{\mathrm{end}}^{2}\frac{4\pi^{4}}{12H^{2}}\frac{\sum_{i}k_{i}^{3}}{\prod_{i}k_{i}^{3}}\mathcal{P}_{+}^{2}N_{l}^{f}N_{m}^{g}N_{n}^{h}\mathcal{T}_{lmn}^{\left(4\right)fgh}\left(\hat{\mathbf{k}}_{1},\hat{\mathbf{k}}_{2},\hat{\mathbf{k}}_{3}\right),\label{eq:Bz1-expr1}\end{equation}
where $g_{\mathrm{end}}\equiv\g/\sqrt{f_{\mathrm{end}}}$ is the strength
of the self-coupling of canonically normalized vector fields at the
end of inflation. Fallowing Refs.~\citep{Karciauskas_etal(2008),Karciauskas_thesis}
we introduce vectors $\mathcal{M}_{i}^{a}$ to simplify expressions
for the bispectrum\begin{equation}
\mathcal{M}_{j}^{a}\left(\mathbf{k}\right)\equiv\mathcal{P}_{+}N_{i}^{a}\te{ij}{}=\mathcal{P}_{+}N_{a}\left[\hat{W}_{j}^{a}-\hat{k}_{j}\left(\hat{\mathbf{W}}^{a}\cdot\hat{\mathbf{k}}\right)\right],\label{eq:M-def}\end{equation}
where no summation over $a$ is assumed. Using this definition and
Eq.~\eqref{eq:T4-def} the bispectrum in Eq.~\eqref{eq:Bz1-expr1}
becomes\begin{equation}
B_{\zeta1}=-4\pi^{4}\frac{\sum_{i}k_{i}^{3}}{\prod_{i}k_{i}^{3}}\frac{g_{\mathrm{end}}^{2}}{12H^{2}}\left(\f{abh}\f{agf}+\f{agh}\f{abf}\right)W_{i}^{b}N_{i}^{g}\left[\mathcal{M}_{j}^{f}\left(\mathbf{k}_{1}\right)\mathcal{M}_{j}^{h}\left(\mathbf{k}_{3}\right)+\mathrm{c.p.}\right].\label{eq:Bz1-final}\end{equation}
In this equation `$\mathrm{c.p.}$' stands for cyclic permutations
of vectors $\mathbf{k}$.

The second contribution to the three-point correlator of the curvature
perturbation is from non-linear terms in Eq.~\eqref{eq:zeta-def}.
As we assume that only vector fields generate $\zeta$ the dominant
term will be \begin{equation}
\left\langle \zeta\left(\mathbf{k}_{1}\right)\zeta\left(\mathbf{k}_{2}\right)\zeta\left(\mathbf{k}_{3}\right)\right\rangle \supset\frac{1}{2}N_{ij}^{ab}N_{m}^{c}N_{n}^{d}\left\langle \delta W_{i}^{a}\star\delta W_{j}^{b}\left(\mathbf{k}_{1}\right)\delta W_{m}^{c}\left(\mathbf{k}_{2}\right)\delta W_{n}^{d}\left(\mathbf{k}_{3}\right)\right\rangle +\mathrm{c.p.}\end{equation}
Using Wicks theorem and Eq.~\eqref{eq:W1-2pnt-def} the bispectrum
from the the above expression becomes\begin{equation}
B_{\zeta2}=4\pi^{4}\frac{1}{k_{1}^{3}k_{3}^{3}}\mathcal{M}_{i}^{a}\left(\mathbf{k}_{1}\right)N_{ij}^{ab}\mathcal{M}_{j}^{b}\left(\mathbf{k}_{3}\right)+\mathrm{c.p.}\label{eq:Bz2-final}\end{equation}

Both $B_{\zeta1}$ and $B_{\zeta2}$ depend not only on the absolute
values of wavevectors $\mathbf{k}$ but also on their direction, making
the bispectrum anisotropic. The total bispectrum is $B_{\zeta}=B_{\zeta1}+B_{\zeta2}$.
To evaluate which term is the dominant one note that $N_{ab}\propto N_{a}/W^{b}$,
where $N_{ab}\equiv\left|N_{ij}^{ab}\right|$. Thus, in order for
the first term to dominate, $g_{\mathrm{end}}W>H$ must be satisfied.%
\footnote{This is in contrast to Refs.~\citep{Bartolo_etal(2009)_Bispectrum,Bartolo_etal(2009)Trispectrum}
where the opposite bound was assumed.%
} If this is the case, the slow-roll condition $\left|\eta_{ij}^{ab}\right|<1$
is violated. Then the evolution of the homogeneous modes of vector
fields becomes strongly non-linear and the above calculations do not
apply. However, if $f$ is modulated by the inflaton or some degree
of freedom which is stabilized at the end of inflation then $f_{\mathrm{end}}$
is equal or very close to unity, i.e. $f_{\mathrm{end}}\lesssim1$
and $\ge\sim\g$. In this case we can expect $\g W\sim H$ if $W$
is not much larger than $H$ (see the footnote on page \pageref{fn:W-g-H-bound}),
as it is natural for $\g$ to be not much bellow unity in particle
physics models. If this is the case, then both contributions to $B_{\zeta}$
are comparable. If, on the other hand, $\ge W<H$ then $B_{\zeta2}$
contribution to the bispectrum dominates.

\section{The End-of-Inflation Scenario\label{sec:Example}}

Let us implement the results of previous sections to a specific example
using the end-of-inflation scenario. This scenario was suggested in
Ref.~\citep{Lyth(2005)EndOfInfl} invoking only scalar fields. In
usual hybrid inflation models inflation ends when the waterfall field
is destabilized by the inflaton. This happens when the inflaton reaches
some critical value and the waterfall field mass becomes tachyonic.
As this critical value is determined solely by the inflaton itself,
inflation ends on a uniform energy density slice. If, as suggested
in Ref.~\citep{Lyth(2005)EndOfInfl}, this critical value is modulated
by some additional field, then the uniform density slice does no longer
coincide with the end-of-inflation slice. This induces the perturbation
in the distance between flat and uniform density slice, which is equal
to the perturbation in $\zeta$. In Ref.~\citep{Yokoyama_Soda(2008)}
it was shown that if the modulating field is $U\left(1\right)$ vector
field, the generated $\zeta$ is in general statistically anisotropic
(see also Ref.~\citep{Karciauskas_etal(2008)}). In this example
we extend the scenario proposed in Ref.~\citep{Yokoyama_Soda(2008)}
to include non-Abelian vector fields.

\subsection{The Model}

Let us consider a Lagrangian which is invariant under transformations
of some non-Abelian symmetry group $G$ \begin{equation}
\mathcal{L}=\frac{1}{2}\partial_{\mu}\varphi\partial^{\mu}\varphi-\frac{1}{4}fF_{a}^{\mu\nu}F_{\mu\nu}^{a}+\frac{1}{2}\mathrm{Tr}\left[\left(D_{\mu}\Phi\right)^{\dagger}D^{\mu}\Phi\right]-V\left(\varphi,\Phi\right),\label{eq:GUT-L}\end{equation}
where $\mathrm{Tr}\left[\ldots\right]$ stands for trace, $\varphi$
is the inflaton field and $F_{\mu\nu}^{a}$ is the field strength
tensor defined in Eq.~\eqref{eq:F-def2}. The gauge kinetic function
$f$ may be the function of the inflaton $f\left(\varphi\right)$.
This has an advantage that we don't introduce additional degrees of
freedom. The behavior of $f\left(\varphi\right)$ for Abelian vector
fields was studied in Refs.~\citep{Kanno_etal(2010)attractor,Dimopoulos_Wagstaff(2010)BackReact}.
It was found that the required scaling $f\propto a^{-4}$ becomes
an attractor solution in a large parameter space. Such kinetic function
was also studied for non-Abelian vector fields in Ref.~\citep{Murata_Soda(2011)nAb-attractor}.
However, for the present purpose we don't need to assume the source
of modulation of $f$.

$\Phi$ in Eq.~\eqref{eq:GUT-L} is the Higgs field corresponding
to a non-trivial representation of $G$ while the covariant derivative
of the Higgs field $D_{\mu}$ is given by\begin{equation}
D_{\mu}=\partial_{\mu}+i\lambda_{A}\mathbf{T}^{a}A_{\mu}^{a},\label{eq:cov-deriv}\end{equation}
where $\mathbf{T}^{a}$ are generators which satisfy the Lie algebra
$\left[\mathbf{T}^{a},\mathbf{T}^{b}\right]=i\f{abc}\mathbf{T}^{c}$
of an unbroken symmetry group $G$ and $\lambda_{A}$ is the gauge
coupling constant, coupling the Higgs field to the vector gauge fields.

The effective potential $V$ in Eq.~\eqref{eq:GUT-L} is taken to
be\begin{equation}
V\left(\varphi,\Phi\right)=\frac{\lambda}{4}\left[\mathrm{Tr}\left(\Phi^{\dagger}\Phi\right)-M^{2}\right]^{2}+\frac{\kappa^{2}}{2}\varphi^{2}\mathrm{Tr}\left(\Phi^{\dagger}\Phi\right)+V\left(\varphi\right),\label{eq:potential}\end{equation}
where $\lambda$, $\kappa$ and $M$ are constants with $M$ being
a symmetry breaking scale. $V\left(\varphi\right)$ is the potential
of $\varphi$ providing the slow-roll inflation. $V\left(\varphi,\Phi\right)$
can be expressed in a more familiar form if we write the Higgs field
as \begin{equation}
\Phi\equiv\phi\mathbf{l},\end{equation}
where $\mathbf{l}$ is the matrix defining the direction of symmetry
breaking in the field space with $\mathrm{Tr}\left[\mathbf{l}^{\dagger}\mathbf{l}\right]=1$.
Then Eq.~\eqref{eq:potential} becomes

\begin{equation}
V\left(\varphi,\phi\right)=\frac{1}{4}\lambda\left(\phi^{2}-M^{2}\right)^{2}+\frac{1}{2}\kappa^{2}\varphi^{2}\phi^{2}+V\left(\varphi\right),\label{eq:V-hybrid}\end{equation}
which is the potential of the hybrid inflation. But in contrast to
the standard hybrid inflation scenarios we assume that the dominant
part of the curvature perturbation is not generated by the inflaton
field. Instead $\zeta$ is generated by the gauge fields through the
gauge coupling constants which couple them to the Higgs field in the
covariant derivative in Eq.~\eqref{eq:cov-deriv}. To see this, note
from Eqs.~\eqref{eq:GUT-L}, \eqref{eq:cov-deriv} and \eqref{eq:V-hybrid}
that the effective mass squared of the Higgs field $\phi$ is

\begin{equation}
m_{\mathrm{eff}}^{2}\left(\mathbf{x}\right)=\kappa^{2}\varphi^{2}-\lambda M^{2}-\lambda_{A}^{2}A_{\mu}^{a}A_{b}^{\mu}\mathbf{l}^{\dagger}\mathbf{T}^{a}\mathbf{T}^{b}\mathbf{l}.\label{eq:m-eff}\end{equation}
In the unitary gauge $\mathbf{l}$ is such that the last term in this
expression is diagonalised to obtain a sum of massive vector fields
$\tilde{M}^{ab}\tilde{A}_{\mu}^{a}\tilde{A}_{b}^{\mu}$. $\tilde{M}^{ab}$
is a diagonal matrix with the only non-zero elements corresponding
to broken generators. Note, however, that $\tilde{M}^{ab}$ is not
the mass matrix of the vector fields but $\phi^{2}\tilde{M}^{ab}$
is. Without the loss of generality, we can arrange generators $\mathbf{T}^{a}$
in such a way that low $a$'s correspond to generators of unbroken
subgroup and higher $a$'s correspond to the broken ones. Then $\tilde{M}^{ab}$
will have non-zero elements only in the lower right block, which can
be written as $M^{\bar{c}\bar{d}}$, where we used bars over indices
to remind us that they run only over the broken generators but not
the full group.

\subsection{The Curvature Perturbation}

The curvature perturbation in this set up has two contributions\begin{equation}
\zeta=\zeta_{\varphi}+\zeta_{\mathrm{e}}.\end{equation}
The first contribution $\zeta_{\varphi}$ is generated at the horizon
crossing during the slow-roll inflation. The resulting power spectrum
of $\zeta_{\varphi}$ is well known to be \citep{Lyth_book} \begin{equation}
\mathcal{P}_{\zeta_{\varphi}}=\frac{1}{2\mpl^{2}\epsilon_{k}}\left(\frac{H_{k}}{2\pi}\right)^{2},\label{eq:Pphi}\end{equation}
where $H_{k}$ and $\epsilon_{k}$ are the Hubble and slow-roll parameters
evaluated at the horizon exit and we used $N_{\varphi}\equiv\partial N/\partial\varphi=\left(2\mpl^{2}\epsilon_{k}\right)^{-1/2}$
for the slow-roll inflation. The contribution of $\zeta_{\varphi}$
to non-Gaussianity is proportional to slow-roll parameters at horizon
exit and, therefore, too small to ever be observable \citep{Maldacena(2002),Seery(2005)snglFdnG}.

When the inflaton crosses some critical value $\varphi_{\mathrm{c}}$,
at which $m_{\mathrm{eff}}^{2}$ in Eq.~\eqref{eq:m-eff} becomes
negative, inflation terminates and the Higgs field rolls down to the
minimum of the potential. From Eq.~\eqref{eq:m-eff} we find \begin{equation}
\varphi_{\mathrm{c}}^{2}\equiv\frac{\lambda}{\kappa^{2}}M^{2}-\frac{1}{\kappa^{2}f}M^{\bar{a}\bar{b}}W_{i}^{\bar{a}}W_{i}^{\bar{b}},\label{eq:phi-c-def}\end{equation}
where we made use of the temporal gauge and Eq.~\eqref{eq:W-def}
to specify $\varphi_{\mathrm{c}}$ in terms of the physical, canonically
normalized vector fields $W_{i}^{\bar{a}}$. The second term in Eq.~\eqref{eq:phi-c-def}
is subdominant, i.e. $\kappa^{2}\jc^{2}\approx\lambda M^{2}$, giving\begin{equation}
\lambda M^{2}\gg M^{\bar{a}\bar{b}}W_{i}^{\bar{a}}W_{i}^{\bar{b}}/f.\label{eq:M-term-dominant}\end{equation}
But, due to perturbations of the vector fields, it modulates the critical
value of the inflaton, making $\varphi_{\mathrm{c}}\left(\mathbf{x}\right)$
a function of space coordinates $\mathbf{x}$. Thus the end of inflation
hypersurface does not coincide with the uniform energy density hypersurface
which results in the generation of the curvature perturbation $\ze$.
Up to the second order $\ze$ is given by \begin{eqnarray}
\ze & = & N_{\mathrm{\e}}\delta\jc+N_{\e\e}\left(\delta\jc\right)^{2},\label{eq:ze-def}\end{eqnarray}
where $N_{\e}\equiv\partial N/\partial\jc=\left(2\mpl^{2}\epsilon_{\e}\right)^{-1/2}$,
$N_{\e\e}\equiv\partial^{2}N/\partial\jc^{2}$ and $\epsilon_{\e}$
is the first slow-roll parameter at the end of inflation. The perturbation
of $\jc$ can be written as %
\footnote{Note that the second term in this equation was neglected in Refs.~\citep{Yokoyama_Soda(2008),Karciauskas_etal(2008)}.%
} \begin{equation}
\delta\jc=\frac{\partial\jc}{\partial W_{i}^{\bar{a}}}\delta W_{i}^{\bar{a}}+\frac{\partial\jc}{\partial f}\delta f,\label{eq:jc-gen}\end{equation}
where $\delta f=\dot{f}\,\delta\jc/\dot{\varphi}_{\mathrm{c}}$ is
the variation of the kinetic function $f$ corresponding to the time
shift from the hypersurface of the uniform energy density to the end
of inflation. As we require $f\propto a^{-4}$ for the gauge and self-couplings
of the canonically normalized fields to be small, the time derivative
of $f$ is negative, $\dot{f}<0$. Thus, the second term in Eq.~\eqref{eq:jc-gen}
suppresses $\ze$. Even more so, if this term dominates $\delta\jc$,
no perturbation is generated at the end of inflation. To ensure, this
does not happen, we require the first term to dominate, which gives
the constraint\begin{equation}
\left(\frac{M^{\bar{a}\bar{b}}W_{i}^{\bar{a}}W_{i}^{\bar{b}}}{\kappa^{2}f_{\e}\jc\mpl}\right)^{2}\ll\epsilon_{\e},\label{eq:1st-term-dom}\end{equation}
where $f_{\e}$ is evaluated just before the end of inflation. To
simplify calculations we assumed a stronger condition, that the second
term in Eq.~\eqref{eq:jc-gen} is completely negligible. Then $\delta\jc$
is equal to\begin{equation}
\delta\jc=-\frac{M^{\bar{a}\bar{b}}W_{i}^{\bar{b}}}{\kappa^{2}f_{\e}\jc}\delta W_{i}^{\bar{a}},\label{eq:jc-result}\end{equation}
and the isotropic part of the power spectrum of $\zeta_{\e}$ in Eq.~\eqref{eq:Pz-iso-def}
is given by

\begin{equation}
\mathcal{P}_{\ze}^{\mathrm{iso}}=\mathcal{P}_{+}\sum_{\bar{a}}N_{\bar{a}}^{2},\label{eq:Piso-end-of-infl-gen}\end{equation}
where from Eqs.~\eqref{eq:ze-def} and \eqref{eq:jc-result}\begin{equation}
N_{i}^{\bar{a}}=-N_{\e}\frac{M^{\bar{a}\bar{b}}W_{i}^{\bar{b}}}{\kappa^{2}f_{\mathrm{e}}\jc}.\label{eq:Na-def}\end{equation}

As was mentioned in subsection~\ref{sub:zeta-Spectrum} if only one
vector field contributes to the curvature perturbation, the anisotropy
in the power spectrum is $\gz=-1$ and such a large value is excluded
by observations. This is the case, for example, with the Abelian vector
field. Thus authors of Refs.~\citep{Yokoyama_Soda(2008),Karciauskas_etal(2008)}
assumed that $\ze<\zeta_{\varphi}$, in which case $\mathcal{P}_{\zeta}^{\mathrm{iso}}$
is dominated by the scalar field contribution and the subdominant
vector field contribution generates anisotropy in the spectrum of
$\zeta$ with $\left|\gz\right|<1$. In our case, since we are dealing
with non-Abelian vector fields, not one but several vector fields
contribute to $\zeta$. If their orientation in space is random, the
anisotropy in the spectrum is suppressed by the number of fields $\mathcal{N}$
(see Eq.~\eqref{eq:Pz-gen}). Thus with the large enough $\mathcal{N}$
(which is evaluated in subsection~\ref{sub:Stat-Anis}) we can generate
the total curvature perturbation without violating observational bounds
on $\gz$. $\mathcal{P}_{\ze}$ dominates the spectrum of the curvature
perturbation if $\left|N_{\bar{a}}\right|\gg N_{\varphi}$. Using
Eq.~\eqref{eq:Na-def} this bound becomes\begin{equation}
\left(\frac{\lambda_{A}^{2}}{f_{\e}}\frac{W}{\kappa^{2}\jc}\right)^{2}\gg\frac{\epsilon_{\e}}{\epsilon_{k}}=\mathrm{e}^{-2N_{\e}\eta},\label{eq:ze-dominates-zphi}\end{equation}
where $\lambda_{A}^{2}/f_{\e}$ is the gauge coupling of the vector
field to the Higgs field at the end of inflation. Evaluating Eq.~\eqref{eq:ze-dominates-zphi}
we assumed that all gauge fields are of the same order, i.e. $W\sim W^{\bar{a}}$
for all $\bar{a}$. We also used the fact that absolute values of
matrix elements of generators $\mathbf{T}^{a}$ in Eq.~\eqref{eq:cov-deriv}
are of order unity so that $\mathrm{Tr}\left(M^{\bar{a}\bar{b}}\right)\sim\lambda_{A}^{2}$.
The slow-roll parameter $\eta$ in Eq.~\eqref{eq:ze-dominates-zphi}
is $\eta\equiv\mpl^{2}V_{\varphi\varphi}\left(\varphi\right)/V\left(\varphi\right)$
and subscripts denote the second derivative of $V\left(\varphi\right)$
with respect $\varphi$. As discussed in Ref.~\citep{Lyth(2005)EndOfInfl}
with $\ze$ dominant, $\eta$ has nothing to do with the spectral
index of the curvature perturbation and can even be $\eta\sim1$.
In this case the right hand side of Eq.~\eqref{eq:ze-dominates-zphi}
can be far bellow unity.

Comparing two bounds in Eq.~\eqref{eq:1st-term-dom} and \eqref{eq:ze-dominates-zphi}
we find that for successful end-of-inflation scenario, in which gauge
fields generate the dominant contribution to the curvature perturbation,
the homogeneous gauge field value must satisfy\begin{equation}
\left(\frac{\mpl}{W}\right)^{2}\epsilon_{\e}\gg\left(\frac{\lambda_{A}^{2}}{\kappa^{2}}\frac{W}{f_{\e}\jc}\right)^{2}\gg\frac{\epsilon_{\e}}{\epsilon_{k}}.\end{equation}
From the first and last terms we find\begin{equation}
\left(\frac{H}{\mpl}\right)^{2}\ll\left(\frac{W}{\mpl}\right)^{2}\ll\epsilon_{k},\label{eq:ze-dom2}\end{equation}
where the first constraint is explained in the footnote on page~\pageref{fn:W-g-H-bound}.
Assuming the scale of inflation to be of the order of GUT scale, i.e.
$\sim10^{16}\,\mathrm{GeV}$, $H/\mpl\sim10^{-4}$. Taking $\epsilon_{k}$,
when cosmological scales leave the horizon, to be of order $10^{-2}$
, the bound in Eq.~\eqref{eq:ze-dom2} gives $10^{-4}\ll W/\mpl\ll10^{-1}$.

\subsection{Anisotropic Spectrum and Bispectrum\label{sub:Stat-Anis}}

To find the full power spectrum of the curvature perturbation generated
by the non-Abelian gauge fields let us substitute Eq.~\eqref{eq:Na-def}
into \eqref{eq:Piso-end-of-infl-gen}. Using Eq.~\eqref{eq:Pz-nonAb}
we find

\begin{equation}
\mathcal{P}_{\zeta}\left(\mathbf{k}\right)=\mathcal{P}_{+}C^{2}\left(M^{2}\right)^{\bar{a}\bar{b}}W^{\bar{a}}W^{\bar{b}}\left[1-\frac{\left(M^{2}\right)^{\bar{a}\bar{b}}\left(\mathbf{W}^{\bar{a}}\cdot\hat{\mathbf{k}}\right)\left(\mathbf{W}^{\bar{b}}\cdot\hat{\mathbf{k}}\right)}{\left(M^{2}\right)^{\bar{a}\bar{b}}W^{\bar{a}}W^{\bar{b}}}\right],\label{eq:Pz-end-of-infl-gen}\end{equation}
where $\left(M^{2}\right)^{\bar{a}\bar{b}}\equiv M^{\bar{a}\bar{c}}M^{\bar{c}\bar{b}}$
is a diagonal matrix, $W^{\bar{a}}$ is the modulus of the vector
field $W^{\bar{a}}\equiv\left|W_{i}^{\bar{a}}\right|$ and $C$ is
defined as \begin{equation}
C\equiv\frac{1}{\kappa^{2}f_{\e}}\frac{N_{\mathrm{\e}}}{\varphi_{\mathrm{c}}}.\end{equation}
Note that $\mathcal{P}_{\ze}$ is determined solely by the massive
vector fields. If the homogeneous values of all vector fields are
of the same order, i.e. $W\sim W^{\bar{a}}$ for all $\bar{a}$, then
the power spectrum in Eq.~\eqref{eq:Pz-end-of-infl-gen} becomes\begin{equation}
\mathcal{P}_{\zeta}\left(\mathbf{k}\right)\approx\lambda_{A}^{4}\mathcal{N}\mathcal{P}_{+}\left(CW\right)^{2}\left[1-\frac{1}{\mathcal{N}}\sum_{\bar{a}}\left(\hat{\mathbf{W}}^{\bar{a}}\cdot\hat{\mathbf{k}}\right)^{2}\right],\label{eq:Pz-end-of-infl}\end{equation}
where $\mathcal{N}$ is the number of \emph{massive} vector fields
and we used $\mathrm{Tr}\left(M^{\bar{a}\bar{b}}\right)\sim\lambda_{A}^{2}$.
As was discussed after Eq.~\eqref{eq:Pz-gen} $\mathcal{N}\geq15$
or $\mathcal{N}\geq4$ is needed to avoid observational constraints
on $\gz$, depending if the systematics causing detected anisotropy
in the spectrum is believed to be the asymmetry of WMAP beams or unknown.
The Lagrangian in Eq.~\eqref{eq:GUT-L} was assumed to be invariant
under the transformation of some non-Abelian symmetry group $G$.
To estimate the minimal rank of the group $G$, which satisfies the
bounds on $\mathcal{N}$, let us assume that $G$ is a special unitary
group $SU\left(N\right)$ which at the phase transition is broken
to $SU\left(N-1\right)$. Such symmetry breaking results in $\mathcal{N}=2N-1$
massive gauge fields. Thus for a weaker bound on $\mathcal{N}$ the
$SU\left(3\right)$ group already generates $g_{\zeta}<0.29$. If,
on the other hand, the stronger bound on $\mathcal{N}$ applies (with
$\left|g_{\zeta}\right|<0.07$), at least $SU\left(8\right)$ is needed.
In realistic particle physics models one has to be careful in choosing
a symmetry group as not to overproduce monopoles after the symmetry
breaking \citep{Jeannerot_etal(2001)GUTinfl}. However, the above
estimate demonstrates that the primordial curvature perturbation can
be generated with gauge fields of reasonably large groups. Even more
so, the anisotropy in the spectrum $g_{\zeta}$, generated by such
groups, are of the magnitude which will be possible to test in the
very near future by the Planck satellite. As it is shown in Refs.~\citep{Pullen_Kamionkowski(2007),Ma_etal(2011)statAnis}
the Planck data will allow to constrain $\gz$ with an accuracy of
$0.01$.

To find the bispectrum for the non-Abelian end-of-inflation scenario
let us differentiate $N_{i}^{\bar{a}}$ in Eq.~\eqref{eq:Na-def}
one more time\begin{equation}
N_{ij}^{\bar{a}\bar{b}}=-\delta_{ij}CM^{\bar{a}\bar{b}}+\frac{N_{i}^{\bar{a}}N_{j}^{\bar{b}}}{\varphi_{\mathrm{c}}N_{\e}}.\label{eq:ddN-end-of-infl}\end{equation}
Using Eq.~\eqref{eq:M-term-dominant} and $\kappa^{2}\jc^{2}\approx\lambda M^{2}$
one can easily check that the first term dominates in this expression.

Also let us find the vector $\mathcal{M}_{i}^{a}\left(\mathbf{k}\right)$
introduced in Eq.~\eqref{eq:M-def}. With $N_{i}^{\bar{a}}$ calculated
in Eq.~\eqref{eq:Na-def}, $\mathcal{M}_{i}^{a}\left(\mathbf{k}\right)$
becomes\begin{equation}
\mathcal{M}_{i}^{\bar{a}}\left(\mathbf{k}\right)=-\mathcal{P}_{+}CM^{\bar{a}\bar{b}}\left[W_{i}^{\bar{b}}-\hat{k}_{i}\left(\mathbf{W}^{\bar{b}}\cdot\hat{\mathbf{k}}\right)\right].\label{eq:M-end-of-infl}\end{equation}

The first bispectrum $B_{\zeta1}$ in Eq.~\eqref{eq:Bz1-final},
which is due to self interactions of gauge fields, can be calculated
using the results in Eqs.~\eqref{eq:Na-def} and \eqref{eq:M-end-of-infl}.
After some algebra we obtain

\begin{eqnarray}
B_{\zeta1} & = & 4\pi^{4}\frac{\sum_{i}k_{i}^{3}}{\prod_{i}k_{i}^{3}}\frac{g_{\mathrm{end}}^{2}}{12H^{2}}C^{3}\mathcal{P}_{+}^{2}\left(\f{ab\bar{h}}\f{a\bar{g}\bar{f}}+\f{a\bar{g}\bar{h}}\f{ab\bar{f}}\right)M^{\bar{g}\bar{c}}M^{\bar{f}\bar{d}}M^{\bar{h}\bar{e}}\left(\mathbf{W}^{b}\cdot\mathbf{W}^{\bar{c}}\right)\times\nonumber \\
 &  & \times\left[\left(\mathbf{W}^{\bar{d}}\cdot\mathbf{W}^{\bar{e}}\right)-2\left(\mathbf{W}^{\bar{d}}\cdot\hat{\mathbf{k}}_{1}\right)\left(\mathbf{W}^{\bar{e}}\cdot\hat{\mathbf{k}}_{1}\right)+\left(\hat{\mathbf{k}}_{1}\cdot\hat{\mathbf{k}}_{3}\right)\left(\mathbf{W}^{\bar{d}}\cdot\hat{\mathbf{k}}_{1}\right)\left(\mathbf{W}^{\bar{e}}\cdot\hat{\mathbf{k}}_{3}\right)+\mathrm{c.p.}\right].\label{eq:Bz1-end-of-infl}\end{eqnarray}
First, note that the bispectrum is anisotropic due to its dependence
on the direction of $\mathbf{k}$ wavevectors. The anisotropy in the
bispectrum from self-interactions is solely determined by massive
vector fields. The amplitude of $B_{\zeta1}$ however is determined
by vector fields and structure constants of the whole group as it
has unbarred indices.

To evaluate the magnitude of $B_{\zeta1}$ let us assume that all
vector fields are of the same order $W$ and $\mathrm{Tr}\left(M^{\bar{a}\bar{b}}\right)\sim\lambda_{A}^{2}$.
Using the isotropic part of the spectrum in Eq.~\eqref{eq:Pz-end-of-infl}
and assuming that structure constants are of order unity we find\begin{equation}
B_{\zeta1}^{\mathrm{iso}}\approx4\pi^{4}\frac{\eta}{2\mathcal{N}}\frac{g_{\mathrm{end}}^{2}W^{2}}{12H^{2}}\left(\mathcal{P}_{\zeta}^{\mathrm{iso}}\right)^{2}\left(\frac{f_{\e}\kappa^{2}\jc^{2}}{\lambda_{A}^{2}W^{2}}\right)\frac{\sum_{i}k_{i}^{3}}{\prod_{i}k_{i}^{3}}.\end{equation}

The expression for the second part of the bispectrum $B_{\zeta2}$
is quite long and the full result is given in Appendix~\ref{sec:Bz2-integral}.
But it is easy to see from Eqs.~\eqref{eq:ddN-end-of-infl}, \eqref{eq:M-end-of-infl}
and \eqref{eq:Bz2-final} that it is anisotropic too and it is determined
solely by massive vector fields. The isotropic part of $B_{\zeta2}$
is\begin{equation}
B_{\zeta2}^{\mathrm{iso}}\approx-4\pi^{4}\frac{\eta}{2\mathcal{N}}\left(\mathcal{P}_{\zeta}^{\mathrm{iso}}\right)^{2}\left(\frac{f_{\e}\kappa^{2}\jc^{2}}{\lambda_{A}^{2}W^{2}}\right)\frac{\sum_{i}k_{i}^{3}}{\prod_{i}k_{i}^{3}}.\end{equation}
Note, that compared to the single field end-of-inflation scenario
in Ref.~\citep{Lyth(2005)EndOfInfl}, the bispectrum is suppressed
by the number of fields. However, as discussed after Eq.~\eqref{eq:Pz-end-of-infl}
we do not expect $\mathcal{N}$ to be too large. Also note, that the
bispectrum from the self-interactions $B_{\zeta1}$ has an additional
factor $\left(g_{\mathrm{end}}W/H\right)^{2}$, where $g_{\mathrm{end}}$
is the strength of self-coupling of the canonically normalized gauge
fields at the end of inflation. Although this factor can not be much
larger than one, as it would make the evolution of gauge fields strongly
non-linear, it might be not much smaller than unity. Finally as is
shown in Eq.~\eqref{eq:Bz2-full} the anisotropy in the bispectrum
as in the spectrum is suppressed by the number of massive gauge vector
fields.

\section{Summary and Conclusions\label{sec:conclusions}}

The possibility of vector fields playing a non-negligible role in
the very early Universe is attracting more and more attention both
from theorists as well as data analists. The role of vector fields
can be to provide either an anisotropically expanding Universe during
inflation or directly affecting or even generating the primordial
curvature perturbation $\zeta$, or both. Some works study the possibility
of vector fields driving inflation too. Because vector field, in contrast
to a scalar field, chooses a preferred direction, the smoking gun
of such models is the statistically anisotropic curvature perturbation.
The effects of such anisotropy can be observed in temperature and
polarization irregularities of the CMB sky. Indeed, with the measurements
of Planck satellite, which is currently collecting data, it will be
possible to constrain statistical anisotropy at the level of $0.01$
\citep{Ma_etal(2011)statAnis}.

In this paper we studied the curvature perturbation generated by non-Abelian
vector fields. Non-Abelian vector fields are one of the main building
blocks of the standard model of particle physics and indeed of any
gauge theory and their existence is an experimentally confirmed fact.
Moreover, theories beyond the standard model contain large numbers
of such fields. In this paper we consider massless non-Abelian fields
with the Lagrangian of the form $\mathcal{L}=-\frac{1}{4}f\left(t\right)F_{a}^{\mu\nu}F_{\mu\nu}^{a}$,
with $F_{\mu\nu}^{a}=\partial_{\mu}A_{\nu}^{a}-\partial_{\nu}A_{\mu}^{a}+\g\f{abc}A_{\mu}^{b}A_{\nu}^{c}$
and $\f{abc}$ being structure constants of a general Lie group. $\g$
is normalized in such a way that $f\left(t_{\mathrm{s}}\right)=1$
when it is stabilized at some time $t_{\mathrm{s}}$. The perturbation
spectrum for Abelian vector fields is flat if the kinetic function
scales as $f\propto a^{-1\pm3}$. In the non-Abelian case however,
$f\propto a^{2}$ corresponds to a strong effective self-coupling
$g\left(t\right)=\g/\sqrt{f\left(t\right)}$ for the physical vector
fields $W_{i}^{a}=\sqrt{f}A_{i}^{a}/a$, which results in strongly
non-linear evolution. For this reason we assume $f\propto a^{-4}$,
which is also necessary for the field perturbation to become classical.
The requirement for predominantly linear evolution of $W_{i}^{a}$
and its perturbation also puts the bound on the self-coupling of fields.
As strong non-linearity affects all scales (not only the ones leaving
the horizon) the bound $\g^{2}<1$ must be satisfied. 

In Refs.~\citep{Kanno_etal(2010)attractor,Dimopoulos_Wagstaff(2010)BackReact}
it was shown that the scaling of the form $f\propto a^{-4}$ can be
achieved dynamically through the backreaction of vector fields on
the evolution of the inflaton, if $f$ is modulated by the inflation
and the vector fields are Abelian. However, this induces anisotropic
expansion of order the slow-roll parameter $\epsilon$. Anisotropic
expansion in its own right introduces an additional source of statistical
anisotropy in the curvature perturbation \citep{Watanabe_etal(2010)anisoBckgr}.
However, in our analysis, we do not require $f$ to be necessarily
modulated by the inflaton. To avoid anisotropic expansion we also
assume a negligible contribution of the vector fields to the the overall
energy budget during inflation.

With this setup in section~\ref{sec:quantum} we calculate the bispectrum
of the field perturbation resulting from interactions of fields. To
calculate the three point correlation function at the tree level we
employ the full quantum perturbation formalism, the so called {}``in-in
formalism''. The interaction Hamiltonian with the above Lagrangian
has two terms, the cubic term with derivative couplings and the quartic
term. The contribution from the first one is suppressed by the physical
momentum $p\ll H$ as compared to the second term. While the bispectrum
from the quartic term is dominated by the classical evolution of fields,
i.e. by interactions after a mode exits the horizon.

This being the case, it is much easier to calculate the correlation
functions from the homogeneous classical equation of motion. Such
calculation is performed in section~\ref{sec:classical}. It is shown
that the result from this method is indeed exactly equal to the dominant
part from the full calculation using the in-in formalism.

In section~\ref{sec:curvature-perturbation} we calculate the spectrum
and bispectrum of the curvature perturbation. It is found that the
spectrum has angular modulation. However, in contrast to the single
vector field case, the anisotropy $g_{\zeta}$ in the spectrum is
suppressed by the number of fields (assuming random orientation).
Thus, reasonably large groups can generate small but observable $\gz$.
In Ref.~\citep{Ma_etal(2011)statAnis} it is shown that with the
Planck data it will be possible to constrain $\gz$ with the precision
up to $0.01$.

The bispectrum of the curvature perturbation has two contributions,
one from the non-Gaussian field perturbation, $B_{\zeta1}$, and the
other from non-linearity in generating $\zeta$, $B_{\zeta2}$. Both
of those contributions have an angular modulation and are comparable
if $g_{\mathrm{end}}W/H\sim1$, where $W$ is the modulus of the homogeneous
part of the vector fields and $g_{\mathrm{end}}$ is the self-coupling
strength of canonically normalized vector fields at the end of inflation.
If this ratio, however, is much larger than 1, the evolution of vector
fields is strongly non-linear and above calculations do not apply.
In the opposite regime $B_{\zeta2}$ dominates.

In the last section~\ref{sec:Example} we present an example of a
mechanism for vector fields to generate $\zeta$. In this example
we consider a scenario in which the curvature perturbation is generated
by varying gauge coupling(s), which couple the Higgs field to vector
bosons in the covariant derivative. In such models $\zeta$ is generated
by the vector boson fields corresponding to broken generators after
the phase transition. We calculate the spectrum and the bispectrum.
The anisotropy in both of them is suppressed by the number of massive
vector gauge bosons.
\begin{acknowledgments}
I would like to thank David H. Lyth, Konstantinos Dimopoulos, Mar
Bastero-Gil and Pere Masjuan for very useful discussions and suggestions.
I am also grateful for the hospitality of CERN Theory Division where
initial stages of this work were completed. This research project
has been partly supported by a Marie Curie Early Stage Research Training
Fellowship of the European Community's Sixth Framework Program under
contract number (MRTN-CT-2006-035863-\textbf{UniverseNet}) it was
also supported by \textbf{CPAN} CSD2007-00042 and \textbf{MICINN}
(FIS2010-17395).
\end{acknowledgments}
\appendix

\section{Calculation of Integrals $I^{\left(4\right)}$ and $I^{\left(3\right)}$
\label{sec:Calculation-of-Intergrals}}

In this appendix we show how to calculate integrals in Eqs.~\eqref{eq:I4-def}
and \eqref{eq:I3-def}. Let us rewrite here the first integral\begin{equation}
I^{\left(4\right)}=\gfi\mathrm{Re}\left[i\int_{\tau_{0}}^{\tau_{\mathrm{end}}}\mathrm{d}\tau'\, a^{8}\left(1-ik_{1}\tau'\right)\left(1-ik_{2}\tau'\right)\left(1-ik_{3}\tau'\right)\mathrm{e}^{ik_{\mathrm{t}}\tau'}\right].\end{equation}
During quasi-de Sitter inflation with $H\approx\mathrm{constant}$
the scale factor is $a\approx-1/\tau H$. Using this and denoting
$x\equiv k_{\mathrm{t}}\tau$, the above integral can be rewritten
as\begin{equation}
I^{\left(4\right)}=\frac{\g^{2}}{f_{0}}k_{\mathrm{t}}^{7}H^{8}\int_{x_{0}}^{x_{\mathrm{end}}}\left(-\frac{\sin x}{x^{8}}+\frac{\cos x}{x^{7}}+K_{1}\frac{\sin x}{x^{6}}-K_{2}\frac{\cos x}{x^{5}}\right)\mathrm{d}x,\label{eq:Int-x}\end{equation}
where $x_{0}\equiv k_{\mathrm{t}}\tau_{0}\rightarrow-\infty$ corresponds
to the initial time when modes are deep within the horizon and $x_{\mathrm{end}}\equiv k_{\mathrm{t}}\tau_{\mathrm{end}}\rightarrow0$
is at the end of inflation. Note that assuming all three $k$'s leave
the horizon at similar time, $N_{k}=-\ln\left(\left|x_{\mathrm{end}}\right|\right)$
is the number of e-folds from the horizon crossing to the end of inflation.
$K_{1}$ and $K_{2}$ are defined in Eq.~\eqref{eq:K1-K2-def}.

The total integral in Eq.~\eqref{eq:Int-x} is the superposition
of integrals $\int\sin x/x^{n}\,\mathrm{d}x$ and $\int\cos x/x^{n}\,\mathrm{dx}$,
with $n$ being a natural number. The order of $n$ within each integral
can be reduced integrating by parts until we arrive at superposition
of terms $\sin x_{\mathrm{end}}/x_{\mathrm{end}}^{n}$ and $\cos x_{\mathrm{end}}/x_{\mathrm{end}}^{n}$
with appropriate constants and the integral $\int\cos x/x\,\mathrm{dx}$.
The last one can be evaluated as follows. Let us write\begin{equation}
\int_{x_{0}}^{x_{\mathrm{end}}}\frac{\cos x}{x}\mathrm{d}x=\int_{-x_{\mathrm{end}}}^{1}\frac{1-\cos x}{x}\mathrm{d}x-\int_{1}^{-x_{0}}\frac{\cos x}{x}\mathrm{d}x-\int_{-x_{\mathrm{end}}}^{1}\mathrm{d}\ln x.\label{eq:cosx-intg}\end{equation}
Taking the limit $x_{\mathrm{end}}\rightarrow-\infty$ and $x_{0}\rightarrow0$
first two terms in Eq.~\eqref{eq:cosx-intg} are equal to the Euler-Mascheroni's
constant $\gamma\approx0.577$ and the last term is $-N_{k}$. Expanding
the result around $x_{\mathrm{end}}\rightarrow0$ and neglecting terms
proportional to $x_{\mathrm{end}}^{n}$ with $n>0$, we arrive at
the final expression in Eq.~\eqref{eq:I4-full}.

The same method can be used to evaluate $I^{\left(3\right)}$. The
difference is, that integrating by parts the lowest order integral
becomes\begin{equation}
\int_{x_{0}}^{x_{\mathrm{end}}}\frac{\sin x}{x}\mathrm{d}x=\frac{\pi}{2},\end{equation}
where we have taken a limit $x_{0}\rightarrow-\infty$ and $x_{\mathrm{end}}\rightarrow0$.

\section{The Bispectrum $B_{\zeta2}$\label{sec:Bz2-integral}}

The expression for the second part of the bispectrum $B_{\zeta2}$
can be calculated using the first, dominant term of $N_{ij}^{\bar{a}\bar{b}}$
in Eq.~\eqref{eq:ddN-end-of-infl}\begin{eqnarray}
\mathcal{M}_{i}^{a}\left(\mathbf{k}_{1}\right)N_{ij}^{ab}\mathcal{M}_{j}^{b}\left(\mathbf{k}_{3}\right) & = & -\mathcal{P}_{+}^{2}C^{3}\left(M^{3}\right)^{\bar{c}\bar{d}}\left[\left(\mathbf{W}^{\bar{c}}\cdot\mathbf{W}^{\bar{d}}\right)-\left(\mathbf{W}^{\bar{c}}\cdot\hat{\mathbf{k}}_{1}\right)\left(\mathbf{W}^{\bar{d}}\cdot\hat{\mathbf{k}}_{1}\right)-\right.\nonumber \\
 &  & \left.-\left(\mathbf{W}^{\bar{c}}\cdot\hat{\mathbf{k}}_{3}\right)\left(\mathbf{W}^{\bar{d}}\cdot\hat{\mathbf{k}}_{3}\right)+\left(\hat{\mathbf{k}}_{1}\cdot\hat{\mathbf{k}}_{3}\right)\left(W^{\bar{c}}\cdot\hat{\mathbf{k}}_{1}\right)\left(W^{\bar{d}}\cdot\hat{\mathbf{k}}_{3}\right)\right],\end{eqnarray}
where $\left(M^{3}\right)^{\bar{a}\bar{d}}\equiv M^{\bar{a}\bar{b}}M^{\bar{b}\bar{c}}M^{\bar{c}\bar{d}}$
is the diagonal matrix. Inserting this result into Eq.~\eqref{eq:Bz2-final}
gives\begin{eqnarray}
B_{\zeta2} & =- & 4\pi^{4}\frac{\left(\mathcal{P}_{\zeta}^{\mathrm{iso}}\right)^{2}}{\mathcal{N}\jc N_{\e}}\left(\frac{\lambda M^{2}}{\lambda_{A}^{2}W^{2}/f_{\e}}\right)\frac{\sum_{i}k_{i}^{3}}{\prod_{i}k_{i}^{3}}\left\{ 1-\frac{1}{\mathcal{N}}\sum_{\bar{a}}\left[k_{2}^{3}\left(\hat{\mathbf{W}}^{\bar{a}}\cdot\hat{\mathbf{k}}_{1}\right)^{2}-k_{2}^{3}\left(\hat{\mathbf{W}}^{\bar{a}}\cdot\hat{\mathbf{k}}_{3}\right)^{2}-\right.\right.\nonumber \\
 &  & \left.\left.-k_{2}^{3}\left(\hat{\mathbf{k}}_{1}\cdot\hat{\mathbf{k}}_{3}\right)\left(\hat{\mathbf{W}}^{\bar{a}}\cdot\hat{\mathbf{k}}_{1}\right)\left(\hat{\mathbf{W}}^{\bar{a}}\cdot\hat{\mathbf{k}}_{3}\right)+\mathrm{c.p.}\right]/\sum_{i}k_{i}^{3}\right\} ,\label{eq:Bz2-full}\end{eqnarray}
where we assumed $W\sim W^{\bar{a}}$ for all $\bar{a}$ and $\mathrm{Tr}\left[\left(M^{3}\right)^{\bar{a}\bar{b}}\right]\sim\lambda_{A}^{6}$.
Note that the anisotropy in the bispectrum is suppressed by the number
of fields, the same suppression as in the spectrum.

\bibliographystyle{aipnum4-1}
\bibliography{Refs_all}

\end{document}